\documentclass{article}

\usepackage{microtype}
\usepackage{graphicx}
\usepackage{subfigure}
\usepackage{booktabs} % for professional tables

\usepackage{hyperref}
\usepackage[dvipsnames]{xcolor}

\usepackage[accepted]{icml2025}

\usepackage{amsfonts}       % blackboard math symbols

\usepackage{microtype}      % microtypography
\usepackage{mathtools}
\usepackage{amsmath, amsthm, amssymb, bm, bbm}
\usepackage{amsmath,amsfonts,bm}

\def\Figref#1{Figure~\ref{#1}}

\def\Secref#1{Section~\ref{#1}}
\def\eqref#1{equation~\ref{#1}}
\def\Eqref#1{Eq.~\ref{#1}}
\def\Appref#1{Appendix~\ref{#1}}

\def\1{\bm{1}}

\newcommand{\normal}{\mathcal{N}}
\newcommand{\bld}[1]{\boldsymbol{#1}}
\def\rvepsilon{\bld{\mathbf{\epsilon}}}

\def\rvg{{\mathbf{g}}}

\def\rvm{{\mathbf{m}}}

\def\rvx{{\mathbf{x}}}
\def\rvy{{\mathbf{y}}}
\def\rvz{{\mathbf{z}}}

\def\rmI{{\mathbf{I}}}

\def\rmS{{\mathbf{S}}}

\DeclareMathAlphabet{\mathsfit}{\encodingdefault}{\sfdefault}{m}{sl}
\SetMathAlphabet{\mathsfit}{bold}{\encodingdefault}{\sfdefault}{bx}{n}

\def\gA{{\mathcal{A}}}

\def\gD{{\mathcal{D}}}
\def\gE{{\mathcal{E}}}

\def\gL{{\mathcal{L}}}

\def\sI{{\mathbb{I}}}

\def\sP{{\mathbb{P}}}

\def\sR{{\mathbb{R}}}

\newcommand{\E}{\mathbb{E}}

\newcommand{\R}{\mathbb{R}}

\DeclareMathOperator*{\argmax}{arg\,max}
\DeclareMathOperator*{\argmin}{arg\,min}

\newcommand{\rvmu}{\bld{\mathbf{\mu}}}

\newcommand{\rmSigma}{\mathbf{\Sigma}}
\newcommand{\rvpsi}{\bld{\mathbf{\psi}}}

\usepackage{multirow}
\usepackage{subfigure}
\usepackage{graphicx}
\usepackage{caption}
\usepackage{soul}
\usepackage{wrapfig}
\usepackage{adjustbox}

\usepackage{tikz}
\usetikzlibrary{arrows.meta, positioning}

\usepackage[capitalize,noabbrev]{cleveref}

\theoremstyle{plain}
\newtheorem{theorem}{Theorem}[section]
\newtheorem{proposition}[theorem]{Proposition}

\theoremstyle{definition}
\newtheorem{definition}[theorem]{Definition}

\theoremstyle{remark}

\usepackage{enumerate}% http://ctan.org/pkg/enumerate

\usepackage[textsize=tiny]{todonotes}

\newcommand{\MN}{LD-SMC}

\newcommand\ignore[1]{{}}

\newcommand\consteq{\stackrel{\mathclap{\normalfont\mbox{c}}}{=}}
\newcommand{\mathleft}{\@fleqntrue\@mathmargin0pt}
\newcommand{\mathcenter}{\@fleqnfalse}

\newcounter{phase}[algorithm]
\newlength{\phaserulewidth}
\newcommand{\setphaserulewidth}{\setlength{\phaserulewidth}}
\setphaserulewidth{.7pt}
\begin{document}
\newcommand{\phase}[1]{%
  % 
  % Top phase rule
  \STATE\leavevmode\llap{\rule{\dimexpr\labelwidth+\labelsep}{\phaserulewidth}}\rule{\linewidth}{\phaserulewidth}
  \vspace{-3ex}
  \STATE\strut\refstepcounter{phase}\textit{Stage~\thephase~--~#1}% Phase text
  % Bottom phase rule
  \vspace{-1.25ex}\STATE\leavevmode\llap{\rule{\dimexpr\labelwidth+\labelsep}{\phaserulewidth}}\rule{\linewidth}{\phaserulewidth}}

\twocolumn[
\icmltitle{Inverse Problem Sampling in Latent Space Using Sequential Monte Carlo}

% It is OKAY to include author information, even for blind
% submissions: the style file will automatically remove it for you
% unless you've provided the [accepted] option to the icml2025
% package.

% List of affiliations: The first argument should be a (short)
% identifier you will use later to specify author affiliations
% Academic affiliations should list Department, University, City, Region, Country
% Industry affiliations should list Company, City, Region, Country

% You can specify symbols, otherwise they are numbered in order.
% Ideally, you should not use this facility. Affiliations will be numbered
% in order of appearance and this is the preferred way.
\icmlsetsymbol{equal}{*}

\begin{icmlauthorlist}
\icmlauthor{Idan Achituve}{Sony}
\icmlauthor{Hai Victor Habi}{Sony}
\icmlauthor{Amir Rosenfeld}{Sony}
\icmlauthor{Arnon Netzer}{Sony}
\icmlauthor{Idit Diamant}{equal,Sony}
\icmlauthor{Ethan Fetaya}{equal,barilan}
\end{icmlauthorlist}

\icmlaffiliation{Sony}{
Sony Semiconductor Israel (SSI), Israel}
\icmlaffiliation{barilan}{Faculty of Engineering, Bar-Ilan University, Israel}
%\icmlaffiliation{sch}{School of ZZZ, Institute of WWW, Location, Country}

\icmlcorrespondingauthor{Idan Achituve}{Idanachi@gmail.com}
%\icmlcorrespondingauthor{Firstname2 Lastname2}{first2.last2@www.uk}

% You may provide any keywords that you
% find helpful for describing your paper; these are used to populate
% the "keywords" metadata in the PDF but will not be shown in the document
\icmlkeywords{Machine Learning, ICML}

\vskip 0.3in
]

% this must go after the closing bracket ] following \twocolumn[ ...

% This command actually creates the footnote in the first column
% listing the affiliations and the copyright notice.
% The command takes one argument, which is text to display at the start of the footnote.
% The \icmlEqualContribution command is standard text for equal contribution.
% Remove it (just {}) if you do not need this facility.

%\printAffiliationsAndNotice{}  % leave blank if no need to mention equal contribution
\printAffiliationsAndNotice{\icmlEqualContribution} % otherwise use the standard text.

\begin{abstract}
In image processing, solving inverse problems is the task of finding plausible reconstructions of an image that was corrupted by some (usually known) degradation operator. Commonly, this process is done using a generative image model that can guide the reconstruction towards solutions that appear natural. The success of diffusion models over the last few years has made them a leading candidate for this task. However, the sequential nature of diffusion models makes this conditional sampling process challenging. Furthermore, since diffusion models are often defined in the latent space of an autoencoder, the encoder-decoder transformations introduce additional difficulties. To address these challenges, we suggest a novel sampling method based on sequential Monte Carlo (SMC) in the latent space of diffusion models. We name our method LD-SMC. We define a generative model for the data using additional auxiliary observations and perform posterior inference with SMC sampling based on a reverse diffusion process. Empirical evaluations on ImageNet and FFHQ show the benefits of LD-SMC over competing methods in various inverse problem tasks and especially in challenging inpainting tasks. 
% In recent years, diffusion models have been utilized as powerful priors for solving inverse problems on image data. The goal in inverse problems is to find plausible reconstructions of an image that was corrupted by some known degradation model. Given the sequential nature of diffusion models, it is sensible to view this task as performing Bayesian inference in a state-space model. The observed variable is the corrupted image, the diffusion model defines a Markovian dynamics, and the aim is to infer a latent quantity (i.e., a clean image). However, when the diffusion process is defined in the latent space of an auto-encoder, tractable Bayesian inference becomes more challenging. Here, we suggest a simple solution based on sequential Monte Carlo (SMC) sampling. We use the forward process of the diffusion model to augment the model with additional observations and then perform posterior sampling by solving an SMC sampling problem. We name our method Gibbs Diffusion Sampling (\MN). \id{need to change Gibbs Diffusion sampling.}
%  We use the forward process of the diffusion model to augment the model with additional observations and then perform posterior sampling by solving an SMC sampling problem. We name our method Gibbs Diffusion Sampling (\MN).
\end{abstract}

\section{Introduction}
\label{Introduction}
Many important signal processing tasks can be viewed as inverse problems \cite{song2021scorebased,Moliner2023audio,Daras2024survey,chungdiffusion,cardoso2023monte}. In inverse problems, the objective is to obtain a clean signal $\rvx \in \R^n$ from a degraded observation $\rvy=\gA(\rvx)+\rvpsi$, where $\gA$ is usually a known irreversible mapping and $\rvpsi$ is a Gaussian noise vector. Common applications that fit this framework include image deblurring, super-resolution, inpainting, and Gaussian denoising. 
The broad applicability of inverse problems makes them highly significant, as they encompass numerous real-world challenges, such as those found in digital image processing \cite{blackledge2005digital}, wireless communication \cite{Chen2021}, seismology \cite{Virieux2009seismology}, medical imaging \cite{song2021scorebased,chung2023solving}, and astronomy \cite{Craig1986book}.
% The wide-ranging applicability of inverse problems underscores their importance, as they address a multitude of practical challenges, including those found in wireless communication, seismology, medical imaging, and astronomy.

%Inverse problems are common in many scientific fields, such as, wireless communication, seismology, medical imaging, and astronomy \ia{[.... Idit - can you please add/alter the text here and add references?]}. 
% In inverse problems, one is given an observation $\rvy \in \R^m$

% a clean signal $\rvx \in \R^n$ undergoes a transformation by a 

% one is given a degraded signal $\rvy$ of some clean, unknown, signal $\rvx$. 

%Many important signal-processing and image-processing tasks can be viewed as inverse problems. 
%In inverse problems, the objective is to retrieve a signal $\rvx\in\R^d$ from a degraded observation $\rvy=\gA(x)+\rvpsi$, where $\gA$ is a known mapping and $\rvpsi$ is a Gaussian noise.

%there is a clean signal $\rvx\in\R^d$ and you are given a corrupted version of it with a known corruption model $\rvy=\gA(x)+\rvpsi$ where $\gA$ is a corruption operator and $\rvpsi$ is a Gaussian noise. 
%\id{suggestion: In inverse problems, the objective is to retrieve a signal $\rvx\in\R^d$ from a degraded observation $\rvy=\gA(x)+\rvpsi$, where $\gA$ is a known mapping and $\rvpsi$ is a Gaussian noise.}

% Common applications that align with this framework encompass tasks such as image deblurring, super-resolution, inpainting, and Gaussian denoising. 

\begin{figure}[!t]
    \centering
    \includegraphics[width=1.0\linewidth]{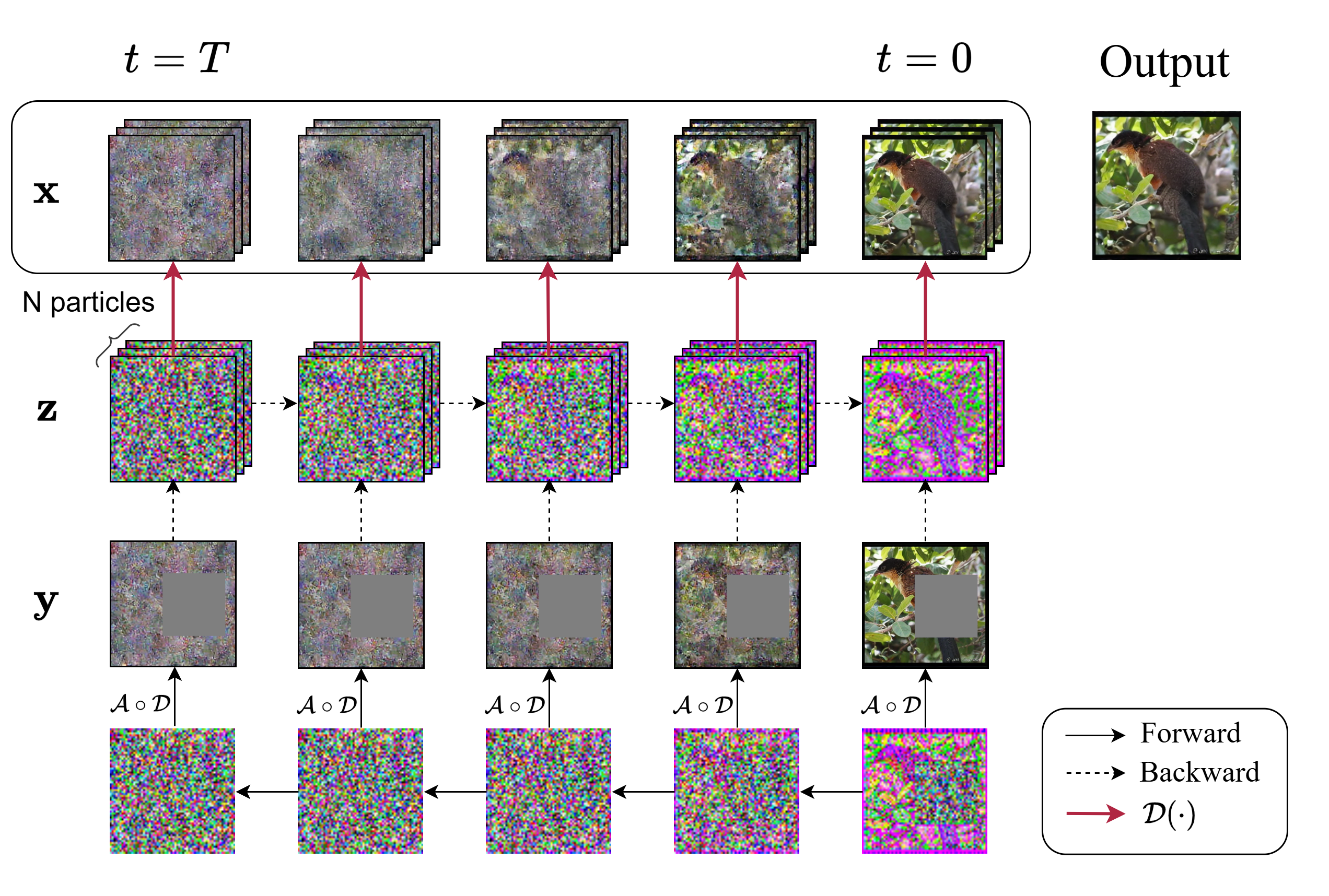}
    \caption{\MN{} solves inverse problem tasks in the latent space of autoencoders by utilizing auxiliary observations $\rvy_{1:T}$ initialized using the DDIM forward process. Then, sampling $\rvz_0$ from the posterior distribution is done based on the reverse diffusion process using sequential Monte Carlo. In the figure, $\gD$ and $\gA$ denote the decoder and the corruption operator respectively.}
    \label{fig:fig1}
\end{figure}

A major challenge in solving inverse problems is the existence of multiple plausible solutions. For example, in image inpainting, the likelihood $p(\rvy|\rvx)$ remains constant regardless of how the absent pixels are filled. However, the desired solution is one that not only fits the observation, but also appears natural, which corresponds to having a high probability under a natural image prior $p(\rvx)$. This insight naturally leads to the approach of sampling from the posterior distribution $p(\rvx|\rvy)\propto p(\rvy|\rvx)p(\rvx)$, combining the data likelihood and the prior to achieve realistic and data-consistent solutions.

%For example, in image inpainting, any way we fill the missing pixels will have the same likelihood $p(\rvy|\rvx)$ value. 
%\id{suggestion: For example, in image in-painting, the likelihood $p(\rvy|\rvx)$ remains constant regardless of how the absent pixels are filled.}

With the impressive recent advances in diffusion models \cite{sohl2015deep,ho2020denoising,SongME21}, there has been a significant interest in leveraging them as prior image models to solve inverse problems. However, integrating diffusion models into this context is not straightforward because of their sequential sampling process. Specifically, diffusion sampling involves iterative drawing from $p(\rvx_{t-1}|\rvx_t)$, while the conditioning on the corrupted image $\rvy$ is defined only in the final step, namely, through $p(\rvy|\rvx_0)$. This mismatch makes direct sampling from the joint posterior $p(\rvx_0,...,\rvx_T|\rvy)$ particularly challenging.

Recently, several studies proposed sequential Monte Carlo (SMC) \cite{doucet2001sequential, del2012adaptive} as an effective solution for this task based on pixel-space diffusion models \cite{cardoso2023monte, trippe2023diffusion, wu2024practical, dou2024diffusion}. Specifically, \citet{wu2024practical} applies the DPS approximation \cite{chungdiffusion} for $p(\rvy|\rvx_t)$ with $p(\rvy|\mathbb{E}[\rvx_0|\rvx_t])$ and uses SMC sampling to correct for it. \citet{dou2024diffusion}, on the other hand, connected $\rvx_t$ to $\rvy$ by introducing a sequence of latent variables $\rvy_{1:T}$ through a duplex forward diffusion process and sampling sequentially from $p(\rvx_{t:T}|\rvy_{t:T})$. While this approach has shown great potential, it has two main limitations. First, it does not take into account future observations $\rvy_{0:t-1}$ in the sampling process. Second, and more crucially, it is limited to linear corruption models only. As such, it cannot be applied with nonlinear mappings $\gA$, let alone common Latent Diffusion Models (LDMs) \cite{rombach2022high} due to the nonlinearity of the decoder. This is a harsh restriction as many of the recent powerful and efficient models are LDMs \cite{esser2024scaling}.

% Since pixel-space diffusion process can be costly, developing models that effectively address inverse problems with latent diffusion models holds significant value. However, compared to pixel-space diffusion models, latent-space diffusion models have been less studied in the context of inverse problems. 

% The latter approach has shown great potential; however, it is limited to linear corruption models only, and as such, cannot be used with common latent diffusion models \cite{rombach2022high} due to the involvement of a non-linear decoder.

%And indeed recently, there have been attempts in that direction (e.g., \cite{song2023solving, rout2024beyond}); however, as we will show in the experimental section, they can often achieve sub-par performance. 

%Furthermore, recently there have been attempts to solve inverse problems using latent diffusion models \cite{song2023solving, rout2024beyond}. 
%We emphasize here that due to the nonlinear decoder in latent diffusion, some previous studies that also used SMC sampling \cite{cardoso2023monte, dou2024diffusion}, cannot be applied as they are suited for problems with a linear corruption model only.

Both existing approaches have pros and cons. Using the $p(\rvy|\mathbb{E}[\rvx_0|\rvx_t])$ approximation can be helpful in capturing the large-scale semantics of the image, but it often lacks in capturing the small details. On the other hand, using auxiliary observations $\rvy_{1:T}$ can help capture finer details, but the duplex forward diffusion process is not amenable to LDMs. Here, we propose a method that combines these two approaches and strives to achieve the best of both worlds. We define a generative model for the data based on a reverse diffusion process, according to which an auxiliary observation $\rvy_t$ is generated directly from $\rvz_t$, the latent diffusion variable. Then, to apply posterior inference over the variables of all timesteps, $\rvz_{0:T}$, we use SMC. To obtain a tractable sampling procedure, we derive an approximate target distributions and define a novel proposal distributions for the SMC sampling process tailored for diffusion models. Hence, we name our method Latent Diffusion Sequential Monte Carlo, or more concisely \MN. An illustration of our approach is shown in \Figref{fig:fig1}. We empirically validated \MN{} on the ImageNet \cite{ILSVRC15} and FFHQ \cite{KarrasLA19} datasets. We found that \MN{} outperforms or is comparable to baseline methods on image deblurring and super-resolution tasks, and can significantly improve over baseline methods on inapainting tasks, especially on the more diverse ImageNet dataset.

In this study, we make the following contributions: (1) we propose a novel method for combining auxiliary observations with latent space diffusion models; (2) we derive approximate target distributions for the SMC procedure and novel proposal distributions to perform approximate posterior sampling specifically tailored for diffusion models; (3) We theoretically show that LD-SMC is asymptotically accurate, namely that it can sample from $p_\theta(\mathbf{z}_0 | \mathbf{y}_0)$; (4) LD-SMC achieves significant improvements in perceptual quality in inpainting tasks, one of the most challenging inverse problem tasks. %Our code is publicly available at \textcolor{green}{\url{https://github.com/IdanAchituve/LD-SMC}} 

\section{Background}
\label{sec:background}
% \textbf{Notations.} Scalars are denoted with lower-case letters (e.g., $x$), vectors with bold lower-case letters (e.g., $\rvx$), and matrices with bold upper-case letters (e.g., $\rmX$).

\textbf{Inverse Problems.} In inverse problems, one would like to recover a sample $\rvx \in \sR^{n}$ from a corrupted version of it $\rvy \in \sR^{m}$. Usually, the corruption model that acted on $\rvx$ is known, but the operation is irreversible \cite{tarantola2005inverse}. For instance, restoring a high-quality image from a low-quality one.
%For instance, cleaning an image from an additive noise sampled from a known noise distribution. 
We denote the corruption operator by $\gA(\cdot)$, %we assume it is governed by a set of parameters $\phi$.
and assume that $\rvy = \gA(\rvx) + \rvpsi$, where $\rvpsi \sim \normal(0, \tau^2\rmI)$ has a known standard deviation $\tau$. In a more concise way, $p(\rvy | \rvx) =\normal(\gA(\rvx), \tau^2\rmI)$. Common examples of inverse problems are inpainting, colorization, and deblurring. In general, solving inverse problem tasks is considered an ill-posed problem with many possible solutions $\rvx$ with equally high $p(\rvy|\rvx)$ values. Given a prior distribution $p(\rvx)$ over natural images, one standard approach to solving the inverse problem is to sample from the posterior distribution $p(\rvx | \rvy) \propto p(\rvy | \rvx) p(\rvx)$.
%Given the ill-posed nature of this problem, an appealing option is to sample from the posterior distribution $p(\rvx | \rvy) \propto p(\rvy | \rvx) p(\rvx)$. Hence, different prior specifications can lead to different recoveries and may have detrimental effect on the quality of the recovered image.

\textbf{Diffusion Models.} Owing to their high-quality generation capabilities, in recent years diffusion models \cite{sohl2015deep, ho2020denoising} have been leveraged as priors in inverse problems \cite{jalal2021robust, songscore}. Here, we adopt the DDIM formulation \cite{SongME21} for the prior model, although our approach can work with other formulations of diffusion models as well. Furthermore, since it is costly to apply the diffusion process in the pixel space, a common approach is to apply the diffusion model in the latent space given by an auto-encoder \cite{rombach2022high}. %Albeit, our method is suited for the pixel space as well.
Applying diffusion models in latent space allows us to sample high-quality images while reducing the computational resources needed by the model. Hence, designing models that effectively solve inverse problems using latent diffusion models is of great importance.

Denote by $\rvz_{0:T}$ the random variables in the latent space.  % In what follows, we use DDPM notations \cite{ho2020denoising}. 
Let $\alpha_{1:T}$, $\beta_{1:T}$ be the variance schedule of the diffusion process with $\beta_t \coloneqq 1 - \alpha_t$. Also, denote by $\bar{\alpha}_t \coloneqq \prod_{j=1}^t \alpha_j$. The DDIM sampling starts from a prior distribution at timestep $T$ set to $p(\rvz_T) = \normal(0, \rmI)$. Then, for $t > 1$ the sampling is done according to $p_{\theta}(\rvz_{t-1} | \rvz_{t}) = \normal(\rvz_{t-1} | \rvmu_\theta(\rvz_t, t), \rmSigma(t))$, where $\theta$ are the parameters of the neural network and, 
\begin{equation}
\begin{aligned} 
    \rmSigma(t) &= \sigma_t^2 \rmI\\
    % \rvmu_\theta(\rvz_t, t) &= \sqrt{\bar{\alpha}_{t-1}}\left(\frac{\rvz_t - \sqrt{1 - \bar{\alpha}_t} \cdot \rvepsilon_\theta (\rvz_t, t)}{\sqrt{\bar{\alpha}_t}}\right) + \\
    \rvmu_\theta(\rvz_t, t) &= \sqrt{\bar{\alpha}_{t-1}}\bar{\rvz}_0(\rvz_t) + \sqrt{1 - \bar{\alpha}_{t-1} - \sigma^2_t} \cdot \rvepsilon_\theta (\rvz_t, t).
\end{aligned}
\label{eq:ddim_sampling}
\end{equation}
Here we denote the approximate posterior mean of $\E[\rvz_0 | \rvz_t]$ by $\bar{\rvz}_0(\rvz_t) \coloneqq \frac{1}{\sqrt{\bar{\alpha}_t}} (\rvz_t - \sqrt{1 - \bar{\alpha}_t} \cdot \rvepsilon_\theta (\rvz_t, t))$ \cite{robbins1956empirical, efron2011tweedie, chungdiffusion}.
For timestep $t=1$, the mean $\rvmu_\theta(\rvz_1, 1)$ is taken to be $\bar{\rvz}_0(\rvz_1)$.
As in \cite{dou2024diffusion} we fix $\sigma_t = \eta \cdot
\sqrt{\beta_t \cdot \frac{1 - \bar{\alpha}_{t-1}}{1 - \bar{\alpha}_t}}$ with $\eta$ being a hyper-parameter.  % = \frac{1}{\sqrt{\bar{\alpha}_t}} (\rvz_t + (1 - \bar{\alpha}_t) \cdot \rvs_\theta (\rvz_t, t))$, where $\rvs_\theta = \nabla_{\rvz_t} \log(p_{\theta}(\rvz_t))$ is an approximation of the score function. 

\textbf{Sequential Monte Carlo (SMC).} SMC is an important technique for sampling in probabilistic graphical models in which exact posterior inference is intractable. The SMC breaks the sampling process down to intermediate steps, allowing efficient sampling through a recursive procedure \cite{doucet2001sequential, del2012adaptive, naesseth2019elements, chopin2020introduction}. 

One family of probabilistic models for which SMC is especially known is state-space models (SSMs), also known as Hidden Markov Models (HMMs). In general, the following quantities must be defined in SSMs, (1) a prior distribution over the initial state $p(\rvz_T)$, (2) a transition distribution that defines the dynamics between states $p(\rvz_t | \rvz_{t+1}) ~ \forall t < T$, and (3) a measurement model $p(\rvy_t | \rvz_t) ~ \forall t < T$. The goal is to sample from the target distribution $p(\rvz_{t:T} | \rvy_{t:T-1})$. To do so, SMC starts by sampling $N$ particles $\{\rvz^{(i)}_T\}_{i=1}^N$ from the prior distribution. Then, at each step, given the previous particle set $\{\rvz^{(i)}_t\}_{i=1}^N$ new samples are taken from a proposal distribution $\rvz^{(i)}_{t-1} \sim \pi_{t-1}(\rvz^{(i)}_{t-1}|\rvz^{(i)}_t)~\forall i \in \{1, ..., N\}$. The particles are then weighted (and possibly resampled) according to the new sequences $\{\rvz^{(i)}_{t-1:T}\}_{i=1}^{N}$.
The proposal distribution serves as an approximation to the target distribution. Its support needs to contain the support of the target density. The weighting mechanism then corrects the approximation by assigning a weight to each particle to adjust its probability. The resampling step, if applied, aims to remove unlikely particles according to the model \cite{sarkka2013bayesian}.
 
%the particles are replaced by new particles that are informed with the data from the current step. Importantly, the sampling process relies on two elements, a proposal distribution and a weighting function. 

%\ef{Given $z^{(i)}_t$ sample from proposal $\pi(z_{t-1}|z^{(i)}_t$)   then given our proposed new sequences $z_{t-1},z^{(i)}_{t:T}$ resample }

\begin{figure}[!t]
\centering
\scalebox{0.8}{
    \begin{tikzpicture}[
        scale=1.0, % Adjust the scale to shrink the entire figure
        node distance=0.4cm and 0.8cm,
        observed/.style={circle, draw, fill=gray!20, minimum size=1.2cm},
        latent/.style={circle, draw, minimum size=1.2cm},
        arrow/.style={-{Stealth[scale=1.2]}}
    ]
    
    % Nodes for observations # , opacity=0
    \node[latent] (yT) {$\rvy_T$};  % , opacity=0
    \node[latent, right=of yT] (yT_1) {$\rvy_{T-1}$};
    \node[right=of yT_1] (dots) {$\cdots$};
    \node[latent, right=of dots] (y1) {$\rvy_1$};
    \node[observed, right=of y1] (y0) {$\rvy_0$};
    
    % Nodes for intermidiate states below observations
    \node[latent, below=of yT] (xT) {$\rvx_T$};  % , opacity=0
    \node[latent, below=of yT_1] (xT_1) {$\rvx_{T-1}$};
    \node[below=of dots, yshift=-0.8cm] (dots2) {$\cdots$};
    \node[latent, below=of y1] (x1) {$\rvx_1$};
    \node[latent, below=of y0] (x0) {$\rvx_0$};
    
    % Nodes for states
    \node[latent, below=of xT] (zT) {$\rvz_T$};
    \node[latent, below=of xT_1] (zT_1) {$\rvz_{T-1}$};
    \node[below=of dots2, yshift=-0.8cm] (dots3) {$\cdots$};
    \node[latent, below=of x1] (z1) {$\rvz_1$};
    \node[latent, below=of x0] (z0) {$\rvz_0$};
    
    % Arrows between states
    % \draw[arrow] (zT_1) -- (zT);
    % \draw[arrow] (dots3) -- (zT_1);
    % \draw[arrow] (z1) -- (dots3);
    % \draw[arrow] (z0) -- (z1);
    % \draw[dashed, arrow] (z1.south) to [out=-60,in=-120] (z0.south);
    % %\draw[dashed, arrow] (dots3.south) to [out=-60,in=-120] (z1.south);
    % \draw[dashed, arrow] (4,-4) to [out=-60,in=-120] (z1.south);
    % %\draw[dashed, arrow] (zT_1.south) to [out=-60,in=-120] (dots3.south);
    % \draw[dashed, arrow] (zT_1.south) to [out=-60,in=-120] (3.5,-4);
    % \draw[dashed, arrow] (zT.south) to [out=-60,in=-120] (zT_1.south);
    \draw[arrow] (zT) -- (zT_1);
    \draw[arrow] (zT_1) -- (dots3);
    \draw[arrow] (dots3) -- (z1);
    \draw[arrow] (z1) -- (z0);

    % Arrows from states to intermidiate states
    \draw[arrow] (z0) -- (x0);
    \draw[arrow] (z1) -- (x1);
    \draw[arrow] (zT_1) -- (xT_1);
    \draw[arrow] (zT) -- (xT);
    
    % Arrows from states to observations
    \draw[arrow] (x0) -- (y0);
    \draw[arrow] (x1) -- (y1);
    \draw[arrow] (xT_1) -- (yT_1);
    \draw[arrow] (xT) -- (yT);
    
    % Add legend for Markovian dynamics
    % \node[below=1cm of x3, align=center] (note) {Graphical model of a state-space model \\ with Markovian dynamics};
    
    \end{tikzpicture}
}
\caption{The graphical model of \MN. In gray observed variables and in white are latent variables.} %We use the forward process of the diffusion model to sample auxiliary observations $\{\rvy_t\}_{t=1}^T$ and then use the reverse process for posterior inference.}
\label{fig:graphical_model}
\end{figure}

\section{Related Work}
\label{sec:related}

Inverse problems have a long and evolving history, with methodologies that have undergone significant advances over the years \cite{Daras2024survey}.
Recently, diffusion models \cite{sohl2015deep, ho2020denoising,songscore} have emerged as effective priors for solving inverse problems in image data \cite{Choi2021ILVRCM,kawar2022denoising,chung2023parallel,chungdiffusion,rout2024solving,song2023pseudoinverse,wangzero,dou2024diffusion,garber2024image,MardaniSKV24,sun2024provable}. 
%We divide our literature review to inverse problem methods in the pixel space and inverse problem methods in the latent space, and note here that several methods can be applied in both settings.
% Another perspective approaches inverse problems from the viewpoint of Bayesian inference. 
% Certain techniques utilize diffusion models as priors to generate plausible reconstructions by sampling from the posterior distribution, for example.

% \textbf{Diffusion-based Inverse Problems in Pixel Space.}
% DDRM \cite{kawar2022denoising} and DDNM \cite{wangzero} utilize diffusion models as prior solving linear inverse problems in pixel space by approximating the measurement matching score, $\nabla \log p(\rvy|\rvx_t)$. 
In \cite{songscore} it was shown that to sample from the posterior distribution, $p(\rvx_0 | \rvy)$, one can solve a stochastic differential equation based on the prior score, $\nabla_{\rvx_t} \log~p_{\theta}(\rvx_t)$, and the conditional score, $\nabla_{\rvx_t} \log~p_{\theta}(\rvy | \rvx_t)$. While the first term is easy to compute, the latter term requires integration over the full diffusion path from time $t$ to $0$. A useful and easy-to-calculate approximation found in several studies is $p_{\theta}(\rvy | \rvx_t) \approx p_{\theta}(\rvy | \E[\rvx_0 | \rvx_t]) \approx p_{\theta}(\rvy | \bar{\rvx}_0(\rvx_t))$, which is readily available at each step \cite{chungdiffusion, song2023pseudoinverse, wu2024practical}.
Specifically, Diffusion Posterior Sampling (DPS) \cite{chungdiffusion} uses this approximation for linear and nonlinear inverse problems with Gaussian and Poisson likelihood models. Other methods also utilize the pseudoinverse of the corruption operator $\gA$ \cite{tirer2018image}. 
%explicit approximations for the measurement matching term with $\mathbb{E}[\rvx_0|\rvx_t]$, 
% approximating $\nabla \log p(\rvy|\rvx_t)$ with $\nabla \log p(\rvy|\mathbb{E}[\rvx_0|\rvx_t])$, 
%addressing non-linear inverse problem scenarios.
$\Pi$GDM \cite{song2023pseudoinverse} introduces a guidance scheme by matching the denoised output and the corrupted image $\rvy$, via a transformation of both. %However, it was observed that relying on that approximation alone may miss fine details in the image \cite{rout2024beyond}.
%  guidance-based approach for inverse problem solving that handles measurements with Gaussian noise, as well as some non-linear, non-differentiable measurement models
DDNM \cite{wangzero} suggested refining the contents of the null space of $\gA$ during the reverse diffusion process. A different approach proposed in \cite{MardaniSKV24} is to approximate the posterior using a variational approach based on the score matching objective presented in \cite{song2021maximum}.
%As such, it is suited only for linear inverse problems. 
%utilizes diffusion models as prior for solving linear inverse problems by decomposing the linear operator $\sA$ into t%in pixel space by approximating the conditional score, $\nabla_{\rvx_t} \log p_{\theta}(\rvy|\rvx_t)$, using $\mathbb{E}[\rvx_0|\rvx_t,\rvy]$.
% Asymptotically Exact Methods
An additional category of inverse problem approaches that use diffusion models is designed with the objective of asymptotic exactness \cite{cardoso2023monte, trippe2023diffusion, wu2024practical, dou2024diffusion}. 
%SMC-Diff \cite{trippe2023diffusion}, MCGDiff \cite{cardoso2023monte}, and TDS \cite{wu2024practical} 
These methods utilize SMC techniques, which have also been applied for unconditional sampling \cite{chen2024sequential}, to target exact sampling from the posterior distribution $p(\rvx_0 | \rvy)$. 
% \ia{for the first two please add why they are not relevant/why we didn't compre to them}.
SMC-Diff \cite{trippe2023diffusion} was designed mainly for motif-scaffolding. It uses the prior distribution as a proposal which is known to require a large number of particles for accurate estimation results \cite{sarkka2013bayesian}, a severe limitation for latent space diffusion models due to expensive decoder evaluations. We experimented with this proposal in our initial attempts and witnessed that even $150$ particles were not enough to sample plausible reconstructions.
% offers asymptotic guarantees solely under the assumption that the trained diffusion model perfectly aligns with the forward noising process, a condition rarely met in practical situations.
MCGDiff \cite{cardoso2023monte} was designed for linear inverse problems only, which makes it unsuitable for inverse problems with latent-space diffusion models due to encoder-decoder involvement. %, with inpainting serving as a specific example.
TDS \cite{wu2024practical}, a recent SMC-based method, solves general inverse problem tasks using the twisting technique. This method also uses the approximation of DPS, but by applying SMC sampling it can correct for it. FPS \cite{dou2024diffusion}, also a recent SMC-based method, uses auxiliary variables. FPS generates a sequence of observations $\rvy_{1:T}$ based on a duplex diffusion process, one process at the $\rvx$ space and the other process at the $\rvy$ space. Since this method is designed for linear inverse problems only, it permits tractable Bayesian inference. Our method combines the ideas of both TDS and FPS to obtain the best of both. Namely, we use the posterior mean approximation and $\rvy_{1:T}$ in our SMC sampling process. As we will show, this combination can be helpful in both understanding the general semantics of an image and capturing fine details. We note here that there are other methods that applied label augmentations similar to FPS, one example is \cite{abu2022adir} which is applicable for linear inverse problems only. Hence, this method as well cannot be trivially combined with latent space diffusion models. 

\begin{algorithm}[!t] 
    \caption{ \MN}
    %\begin{algorithmic}
        1: Set $\rvz_0 = \gE(\rvx_0)$, where $\rvx_0 = \argmin_\rvx || \rvy_0 - \gA(\rvx)||^2_2$\\
        2: Sample $\rvz_{1:T} \sim  p(\rvz_{1:T} | \rvz_0)$ according to the forward \\ 
        \hspace*{3mm} process of DDIM (\Eqref{eq:ddim_forward}) \\
        3: {\bf For} $k = 1, ..., K$:\\
        4: \quad Sample $\rvy_{1:T} \sim \prod_{t=1}^{T} \normal(\rvy_t | \gA(\gD(\rvz_t)), \tau^2\rmI)$\\
        5: \quad Sample $\rvz_{0:T} \sim p_{\theta}(\rvz_{0:T}|\rvy_{0:T})$ using $\mathbf{SISR}(\rvy_{0:T})$\\
    
        6: $\mathbf{SISR}(\rvy_{0:T})$:\\
        7: \quad \hspace*{0.5mm} {\bf For} $i = 1, ..., N$:\\
        8: \quad \quad Sample $\rvz_T^{(i)} \sim p_\theta(\rvz_T)$ and initialize $\rvz^{(i)} = (\rvz_T^{(i)})$\\ 
        % 9: \quad \quad Set, $w_T^{(i)} \propto p_\theta(\rvy_T | \rvz_T) \bar{p}_\theta(\rvy_0 | \rvz_T)$ and normalize \\
        % \hspace*{10mm} the weights to sum to one.\\
        9: \quad \quad Set $w_T^{(i)}$ according to \Eqref{eq:weight_functions}\\
        10: \quad Normalize $\{w_{T}^{(i)}\}_{i=1}^{N}$ to sum to one\\
        11: \quad {\bf For} $t = T-1, ..., 0$:\\
        12: \quad \quad $\{\rvz^{(i)}\}_{i=1}^{N} \sim \text{Multi}(\{\rvz^{(i)}\}_{i=1}^N, \{w_{t+1}^{(i)}\}_{i=1}^N)$,\\
        \hspace*{12mm} a resampling step\\
        13: \quad \quad {\bf For} $i = 1, ..., N$:\\
        % 14: \quad \quad \quad Sample from proposal distribution (\Eqref{eq:proposal}), \\ 
        % \hspace*{15mm} $\rvz_{t}^{(i)} \sim \pi_t(\rvz_{t} | \rvz_{t+1}^{(i)})$\\
        14: \quad \quad \quad Sample $\rvz_{t}^{(i)} \sim \pi_t(\rvz_{t} | \rvz_{t+1}^{(i)})$ (\Eqref{eq:proposal}) \textcolor{OliveGreen}{\# proposal} \\
        15: \quad \quad \quad Set $w_{t}^{(i)}$ according to \Eqref{eq:weight_functions}\\
        16: \quad \quad \quad Update $\rvz^{(i)} = (\rvz_{t}^{(i)}, ..., \rvz_T^{(i)})$\\
        17: \quad \quad Normalize $\{w_{t}^{(i)}\}_{i=1}^{N}$ to sum to one\\
        % 17: \quad \quad Concatenate $\rvz^{(i)} = (\rvz_{t}^{(i)}, ..., \rvz_T^{(i)}) \quad i = 1, ..., N$\\
        18: \quad Sample one chain $\rvz \sim \text{Multi}(\{\rvz^{(i)}\}_{i=1}^N, \{w_{0}^{(i)}\}_{i=1}^N)$\\
        19: \quad {\bf Return} $\rvz$
    \label{algo:LD_SMC}
\end{algorithm}

\begin{figure*}[!t]
    \centering
    \includegraphics[width=0.72\linewidth]{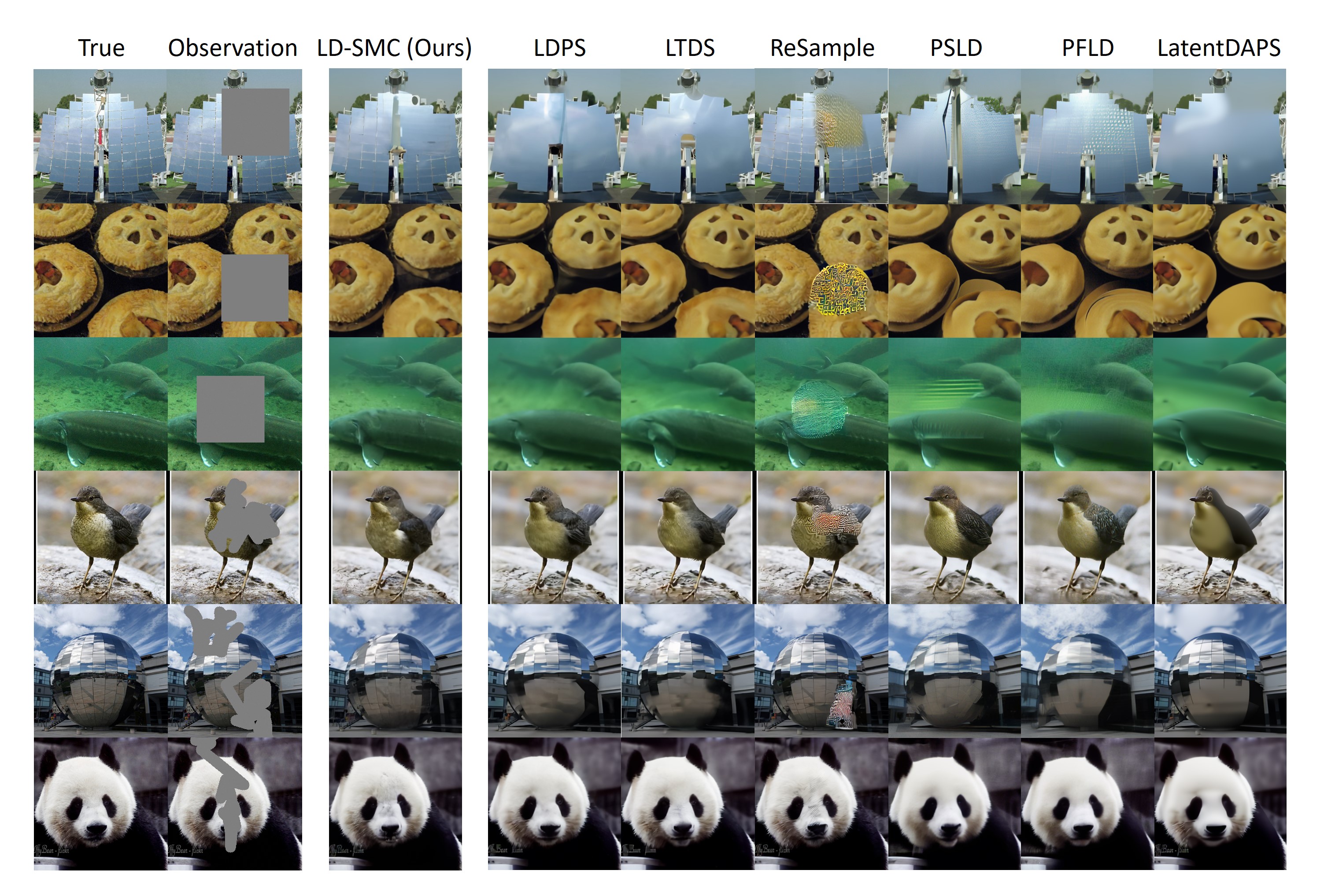}
\caption{Comparison between \MN{} and baseline methods on inpainting of ImageNet images.}
\label{fig:in_ip_recon}
\end{figure*}

%Importantly, the connection to the observation is made only through $\E[\rvx_0 | \rvx_t]$. 

% demonstrated considerable potential linking posterior sampling with Bayesian filtering and adeptly tackling the filtering problem utilizing sequential Monte Carlo methods. However, this approach is confined to linear corruption models. As we will show in \Secref{sec:method} our method combines TDS and FPS to solve general inverse problems.
% The FPS methodology \cite{dou2024diffusion} links posterior sampling with Bayesian filtering and adeptly tackles the filtering problem utilizing sequential Monte Carlo methods. This approach demonstrates considerable potential, but it is confined to linear corruption models.

% Recently, several approaches \cite{dou2024diffusion,wu2024practical,cardoso2023monte} have modeled inverse problems as Bayesian inference in a state-space model.
% However, these methods were designed for diffusion models in pixel space.
% and with the exception of \cite{cardoso2023monte} they assume a linear degradation model.

Several inverse sampling methods were specifically tailored for latent diffusion models. 
PSLD \cite{rout2024solving} extend DPS \cite{chungdiffusion} by incorporating an additional gradient update step to guide the diffusion process to sample latent representations that maintain the integrity of the decoding-encoding transformation, ensuring it remains non-lossy.
% for which the decoding-encoding map is not lossy.
STSL \cite{rout2024beyond} uses second-order information for reconstruction based on efficient approximations.  Comparative analysis with STSL was not feasible due to the absence of publicly available code, making replication challenging.
% a Second-order Tweedie sampler from Surrogate Loss.  
ReSample \cite{song2023solving}, a contemporary method alongside PSLD, acts in two stages; it first applies hard data consistency to obtain latent variables that are consistent with the observed measurements and then employs a resampling scheme to map the sample back onto the data manifold. 
Concurrent to this study \citet{nazemi2024particle} proposed a particle filtering approach. Their method builds on PSLD and DPS update in the proposal distribution. Similarly to TDS \cite{wu2024practical} the connection to the labels is only through $\rvz_0$ using the approximate mean estimator. Also, unlike LD-SMC it does not enjoy asymptotic guarantees. Recently, both DAPS \cite{zhang2024improving} and MGPS \cite{moufad2024variational} have been proposed to address inverse problems both in pixel and latent space. DAPS decouples latents in the sampling trajectory by first sampling from $\rvz_0 | \rvz_{t + \Delta t}, \rvy$ using MCMC techniques, for some $\Delta t > 0$, and then sampling $\rvz_t$. MGPS also decomposes the sampling, but to an intermediate midpoint and uses variational inference for posterior sampling. Given the similarities of both methods, in our experiments we compare only to the former. SILO \cite{raphaeli2025silo}, an additional recent study, first learns a NN that mimics an encoder operation on labels $\rvy$ which is then used to guide the sampling process. In our empirical evaluation, we compare to methods that apply sampling only.

% refined based on a proposal distribution and a weighting function. The proposal distribution serves as an approximation to the posterior distribution and should uphold two conditions: (1) its support contains the support of the posterior density and (2) it is easy to sample from. The weighting function corrects the approximation by assigning a weight for each particle.
%Let $\pi(\rvz_t | \rvz_{t+1})$ denote the proposal density at timestep $t$

%which corresponds to a Markovian forward process as in \cite{dou2024diffusion} 

% \textbf{Particle Filtering.} Particle filtering is a technique to sample from a posterior distribution in state space models \cite{sarkka2013bayesian}. The key idea 

\section{Method}
\label{sec:method}
%We first provide an overview of our approach and then we will describe each element of it in detail. 
Given a corrupted image $\rvy_0$, the goal is to sample $\rvz_0 \sim p_{\theta}(\rvz_0 | \rvy_0)$ using a pre-trained latent diffusion model as prior. Then, we can transform this sample into an image by applying a pre-trained decoder $\gD$, i.e. $\rvx_0 \coloneqq \gD(\rvz_0)$. Here, we stack all the parameters of the networks (diffusion and encoder-decoder) under $\theta$. In what follows, we first define a generative model for the data and then describe how to perform Bayesian inference on all latent variables using blocked Gibbs sampling. Specifically, we use the diffusion process and the corruption operator to augment the model with additional auxiliary observations. Then, inference is applied over the diffusion variables using SMC. The corresponding graphical model can be seen in \Figref{fig:graphical_model}. %Due to the iterative nature of Gibbs sampling, this procedure can be applied multiple times, yet in our experiments, we found that one forward-backward pass suffices to achieve good results.
\subsection{The Generative Model}
\label{method:gen_model}
We now explicitly define the data generation model,
%We add an observation for Augmenting the model with additional random variables $\rvz_{1:T} \coloneqq \{\rvz_t\}_{t=1}^{T}$ and $\rvy_{1:T-1} \coloneqq \{\rvy_t\}_{t=1}^{T-1}$ and define the following generative process:
\begin{equation*}
\begin{aligned}
    % &1.~ \rvz_0 \sim p(\rvz_0)\\
    % &2.~ \rvz_T | \rvz_0 \sim \normal(\sqrt{\bar{\alpha}_T}\rvz_0, (1 - \bar{\alpha}_T) \rmI)),\\
    % &3.~ \rvz_{t-1} | \rvz_{t}, \rvz_0 \sim p(\rvz_{t-1} | \rvz_{t}, \rvz_0) \quad \forall t \in \{2, ..., T\},\\
    % &4.~ \rvy_t | \rvz_t \sim \normal( \gA(\underbrace{\gD(\rvz_t)}_{\rvx_t}), \tau^2\rmI) \quad \forall t \in \{0, ..., {T}\}.
    &1.~ \rvz_T \sim \normal(0, \rmI)),\\
    &2.~ \rvz_{t-1} | \rvz_{t} \sim p(\rvz_{t-1} | \rvz_{t}) \quad \forall t \in \{1, ..., T\},\\
    &3.~ \rvy_t | \rvz_t \sim \normal( \gA(\underbrace{\gD(\rvz_t)}_{\rvx_t}), \tau^2\rmI) \quad \forall t \in \{0, ..., {T}\}.
    % &4. \hat{\rvz}_T \sim \normal(0, \rmI)\\
    % &5.~ \hat{\rvz}_t | \hat{\rvz}_{t+1}, \rvy_t \sim p_\theta(\rvz_t | \hat{\rvz}_{t+1}, \rvy_t) \quad \forall t \in \{{T-1}, ..., 0\}\\
    % &6.~ \rvy_0 | \hat{\rvz}_0 \sim \normal(\rmA\gD(\hat{\rvz}_0), \sigma^2\rmI)
\end{aligned}
\label{gen_proces}
\end{equation*}
Where, $p_\theta(\rvy_t | \rvz_t)$ is defined by the corruption model,
$p(\rvz_{t-1} | \rvz_{t}) = \E_{p(\rvz_0|\rvz_t)}[p(\rvz_{t-1} | \rvz_{t}, \rvz_0)]$ is a backwards generative process that corresponds to a non-Markovian forward process with
\small
\begin{equation}
\begin{aligned}
    &p(\rvz_{t-1} | \rvz_{t}, \rvz_0) =  \\
    &\normal \left( \rvz_{t-1}| \sqrt{\bar{\alpha}_{t-1}}\rvz_0 + \sqrt{1 - \bar{\alpha}_{t-1} - \sigma^2_{t}} \cdot \frac{\rvz_{t} - \sqrt{ \bar{\alpha}_{t}} \rvz_0}{\sqrt{1 - \bar{\alpha}_t}}, \sigma_t^2 \rmI \right),
\end{aligned}
\label{eq:ddim_forward}
\end{equation}
\normalsize
set according DDIM \cite{SongME21}. 
Introducing unobserved data is a known technique in statistics for conducting effective Markov chain Monte Carlo (MCMC) sampling \cite{van2001art, dou2024diffusion}. In our case, we can use it while leveraging the dependencies between the variables in order to build an efficient SMC sampling procedure, as described in the next section.
%Note that we intentionally write the forward diffusion with dependence between subsequent timesteps (instead of $\rvz_0$), it will be made clear why in a few moments. 
%, and $p_\theta(\rvz_t|\rvz_{t+1}, \rvy_t)$ models the dynamics.

\subsection{Sampling Procedure}
\label{sec:sampling_procedure}
Given the generative model defined in \Secref{method:gen_model}, our aim is to apply Bayesian inference over the latent variables. In broad strokes, to obtain a sample $\rvz_0 \sim p_{\theta}(\rvz_0 | \rvy_0)$ we use blocked Gibbs sampling to sample in turns from $p_\theta(\rvy_{1:T}|\rvz_{0:T},\rvy_{0})$ using knowledge on the corruption model and $p_{\theta}(\rvz_{0:T}|\rvy_{0:T})$ using SMC. Specifically, we propose the following procedure:
\begin{enumerate}
    \item Obtain an initial guess for $\rvz_0$ (Sec. \ref{sec:init}).
    \item Sample, $\rvz_{1:T} \sim p(\rvz_{1:T} | \rvz_0, \rvy_0) = p(\rvz_{1:T} | \rvz_0)$ according to the forward process of DDIM (\Eqref{eq:ddim_forward}).
    \item Repeat for some fixed number of steps:
    \begin{enumerate}
        \item Sample, $\rvy_{1:T} \sim p_{\theta}(\rvy_{1:T}|\rvz_{0:T}, \rvy_0) = p_{\theta}(\rvy_{1:T}|\rvz_{1:T}) = \prod_{t=1}^{T} \normal(\rvy_t | \gA(\gD(\rvz_t)), \tau^2\rmI)$.
        \item Sample, $\rvz_{0:T} \sim p_{\theta}(\rvz_{0:T}|\rvy_{0:T})$ using SMC based on a pre-trained diffusion model (Sec. \ref{sec:post_sampling}).
    \end{enumerate}
\end{enumerate}
% $p(\rvz_{1:T}|\rvz_0,\rvy_{0:T-1})=p(\rvz_{1:T}|\rvz_0)$, the auxiliary observations $p(\rvy_{1:T-1}|\rvz_{0:T},\rvy_0)$, and the backward diffusion variables $p({\rvz}_{0:T}|\rvy_{0:T-1})$.
% from $p_{\theta}(\rvz_{1:T}, \rvy_{1:T-1}, \hat{\rvz}_{0:T-1} | \rvy_0)$. Then, we can take samples of $\hat{\rvz}_0$ only. 
Here, we use the dynamics of the forward process and the graphical model dependencies in steps 2 and 3(a). The two steps that are not straightforward are obtaining an initial guess for $\rvz_0$ (step 1) and performing the sampling process in step 3(b). We discuss both in the following sections, but before that we make a few comments. First, clearly the run-time of the algorithm depends linearly on the number of Gibbs iterations. As a result, the sampling time can be slow. However, with proper initialization (as discussed in \Secref{sec:init}), we empirically found that even one iteration of Step 3 suffices to achieve good results. Hence, unless otherwise stated, in our empirical evaluations, we applied this step only once. Second, one may be concerned that evaluating the decoder on noisy latent variables can generate non-natural images. While this may be true, empirically we observed a graceful degradation in image quality with time, as seen in \Figref{fig:in_forward_process}. Furthermore, the latent variables, even if noisy, carry information about $\rvy_0$. This information is then transferred to the auxiliary labels which help guide the sampling process. %Third, we use DDIM in step 3(b), in the 

%from the forward process having a small std $\xi$. In our experiments, we found that \MN{} was robust to the std choice; however, having a small std helped to obtain a higher likelihood in the early stages of the backward sampling procedure. 

%, but before that we would like to make two comments regarding the sampling procedure. First, in our experiments, we found that one forward-backward pass suffices to achieve good results. Hence, unless otherwise stated, we apply Step 2 only once. Second, the backward process in step (d) starts with $\rvz_T$ that was obtained from the forward process. 
%We note here that in practice in step (a) we used the DDIM sampler. \ia{should we say that here? Is it relevant?}\ef{then write the DDIM equation in step 1a}
% since it allowed us to control the variance of the forward process as well using the hyper-parameter $\eta$. 

\begin{figure*}[!t]
    \centering
    \includegraphics[width=0.72\linewidth]{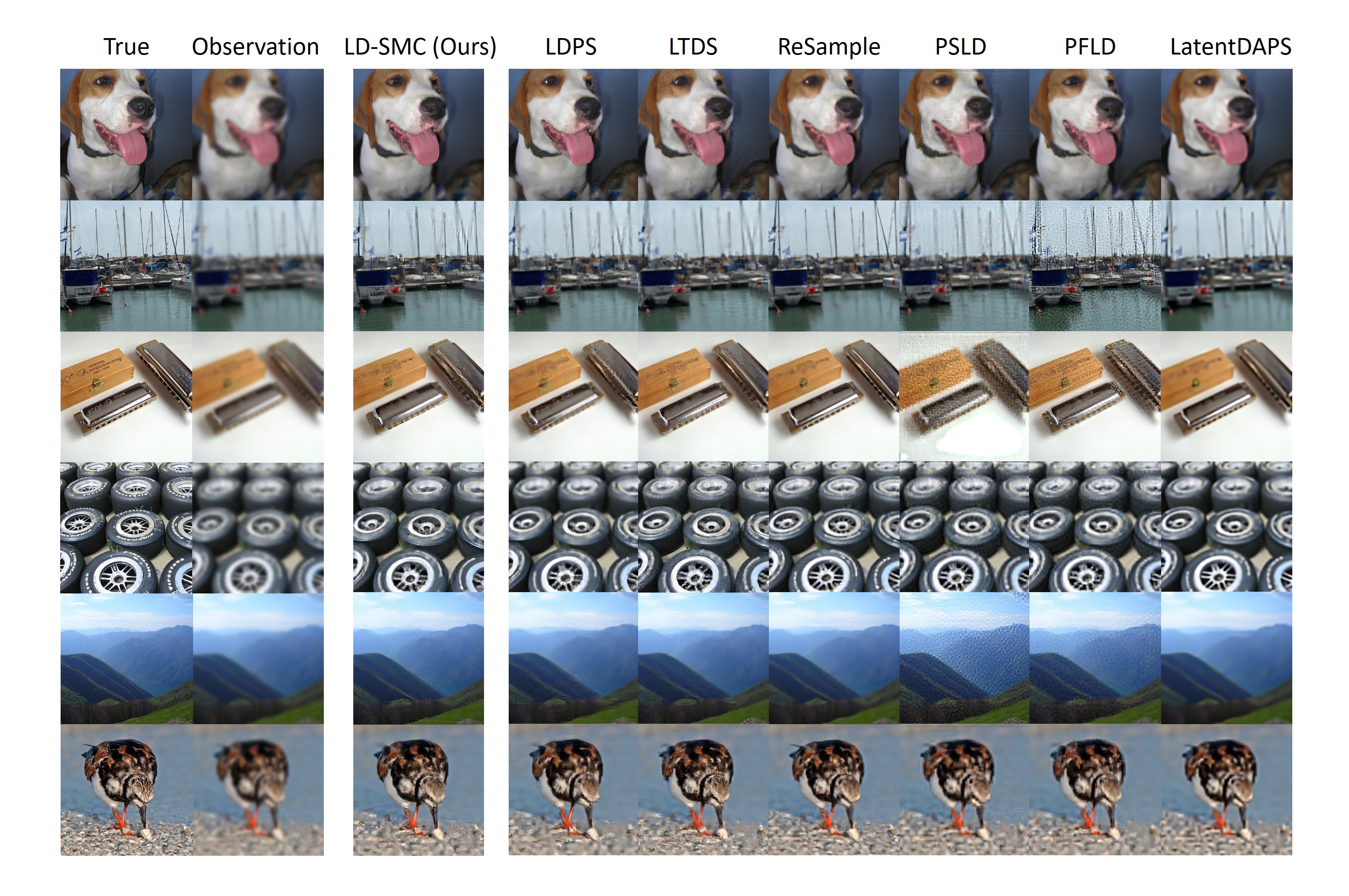}
\caption{Comparison between \MN{} and baseline methods on Gaussian deblurring of ImageNet images.}
\label{fig:ffhq_gd_recon_comp}
\end{figure*}

\subsubsection{Initial Guess for $\rvz_0$}
\label{sec:init}
The first challenge is to obtain an initial $\rvz_0$ (step 1. in the sampling procedure). There are multiple sensible ways to perform this initialization. Importantly, $\rvz_0$ should carry information about the measurement $\mathbf{y}_0$, which can then be used to guide the sampling process using the auxiliary labels. As a result, the variance in the process will be reduced and the convergence will be accelerated. For example, one option is to leverage the pseudoinverse of $\gA$ for linear operators similar to \cite{wangzero}. But, that would not work for nonlinear corruption operators. Although there are mitigations for this problem \cite{song2023pseudoinverse}, to maintain generality, we applied the following optimization procedure, 
% (1) Solve the following optimization problem in latent space:
% \begin{equation}
% \begin{aligned}
%     \hat{\rvz}_0 = \argmax_\rvz log~p_{\theta}(\rvy_0 | \rvz) = \argmin_\rvz || \rvy_0 - \gA(\gD(\rvz))||^2_2
% \end{aligned}
% \label{optimize_z}
% \end{equation}
% (2) Solve the following optimization problem in pixel space and then apply the encoder on it:
\begin{equation}
\begin{aligned}
    \rvx_0 &= \argmax_\rvx log~p_{\theta}(\rvy_0 | \rvx) = \argmin_\rvx || \rvy_0 - \gA(\rvx)||^2_2,
\end{aligned}
\label{optimize_x}
\end{equation}
where $\rvx$ is initialized to $\gD(\hat{\rvz}_0)$ for $\hat{\rvz}_0 \sim \normal(0, \rmI)$.
Then, after reaching convergence, an initial $\rvz_0$ is obtained by applying the encoder on the outcome, namely $\rvz_0 = \gE(\rvx_0)$.
An alternative for this procedure is to perform the optimization process directly in the latent space. However, in our experiments we found that the former option worked better and was substantially faster as it did not involve expensive gradient propagation through the decoder.

\subsubsection{Posterior Sampling}
\label{sec:post_sampling}
We now move on to explain step 3(b) of the sampling procedure.
Due to the non-linearity of the decoder, even for linear corruption operators $\gA$, finding the exact posterior is intractable. One option to overcome this difficulty is to use SMC sampling. In what follows, we first describe approximate (tractable) target distributions for $\{p_{\theta}(\rvz_{t:T}|\rvy_{0:T})\}_{t=0}^{t=T}$, then we present the induced sequence weights $w_t$ and a novel proposal distributions $\pi_t$ for all timesteps $t$. These components are used as part of a general sequential importance sampling with resampling (SISR) algorithm as depicted in Algorithm \ref{algo:LD_SMC}.

%For notational convenience we drop the dependence in $\{\hat{\rvz}_T^{(i)}\}_{i=1}^{N}$ in $p_\theta$, as it is only relevant at the initial SMC step.

\textbf{Approximate target distributions.} First, notice that due to the structure of the model, the posterior density of the r.v. $\rvz_t$ at each step $t$ depends only on $\rvz_{t+1:T}$. Hence, the target distribution of interest at each timestep $t$ is $p_{\theta}(\rvz_{t:T}| \rvy_{0:T})$. However, even computing the unnormalized quantity of that target distribution can be costly. Therefore, we make the following assumption $p_{\theta}(\rvz_{t:T}| \rvy_{0:T}) \approx p_{\theta}(\rvz_{t:T}| \rvy_{t:T}, \rvy_{0})$. This assumption is reasonable since $\rvy_0$ stores all the input information to begin with. In Appendix \ref{sec:approx_target_dist} we arrive at the following recursive formula which forms the approximate target distributions of the SMC procedure for all $t$:
% \[
% \begin{cases}
%     p_\theta(\rvz_{T} | \rvy_{T}, \rvy_0) \propto p_\theta(\rvy_T | \rvz_{T}) \bar{p}_\theta(\rvy_0 | \rvz_{T})p(\rvz_{T}) , \quad t = T\\
%      p_\theta(\rvz_{t:T} | \rvy_{t:T}, \rvy_0) \propto \qquad \qquad \qquad  \qquad ~~~~~~ 0 < t < T \\
%      \frac{p_\theta(\rvy_t | \rvz_t) \bar{p}_\theta(\rvy_0 | \rvz_t)}{\bar{p}_\theta(\rvy_0 | \rvz_{t+1})}p_\theta(\rvz_t|\rvz_{t+1})p_\theta(\rvz_{t+1:T} | \rvy_{t+1:T}, \rvy_0), \\
%       % p_\theta(\rvz_{0:T} | \rvy_{0:T}) \propto \frac{p_\theta(\rvy_0 |\rvz_0)}{\bar{p}_\theta(\rvy_0 |\rvz_1)}p_\theta(\rvz_0 |\rvz_1) p_\theta(\rvz_{1:T} | \rvy_{0:T}), t = 0. \\
%      p_\theta(\rvz_{0:T} | \rvy_{0:T}) \propto \qquad \qquad \quad \qquad \qquad \qquad \quad ~~~~ t = 0 \\
%      \frac{p_\theta(\rvy_0 |\rvz_0)}{\bar{p}_\theta(\rvy_0 |\rvz_1)}p_\theta(\rvz_0 |\rvz_1) p_\theta(\rvz_{1:T} | \rvy_{0:T}).
% \end{cases}
% \]

\begin{equation}
\begin{aligned}
     &p_\theta(\rvz_{t:T} | \rvy_{t:T}, \rvy_0) \propto \\
    &\begin{cases}
    p_\theta(\rvy_T | \rvz_{T}) \bar{p}_\theta(\rvy_0 | \rvz_{T})p(\rvz_{T}) , \qquad \qquad \qquad \qquad  t = T,\\
     \frac{p_\theta(\rvy_t | \rvz_t) \bar{p}_\theta(\rvy_0 | \rvz_t)}{\bar{p}_\theta(\rvy_0 | \rvz_{t+1})}p_\theta(\rvz_t|\rvz_{t+1})p_\theta(\rvz_{t+1:T} | \rvy_{t+1:T}, \rvy_0), \\
    \qquad \qquad \qquad  \qquad \qquad \qquad \qquad \qquad \quad ~~ 0 < t < T,\\
     \frac{p_\theta(\rvy_0 |\rvz_0)}{\bar{p}_\theta(\rvy_0 |\rvz_1)}p_\theta(\rvz_0 |\rvz_1) p_\theta(\rvz_{1:T} | \rvy_{0:T}), \qquad \qquad \qquad t = 0,
\end{cases}
\end{aligned}
\label{eq:approx_target}
\end{equation}

%, and $\rvy_{0:t}$ $p_{\theta}(\rvz_{0:T}|\rvy_{0:T-1})$  
% can be factorized according to,
% \begin{equation*}
%     \begin{aligned} 
%         &p_{\theta}(\rvz_{0:T}|\rvy_{0:T-1}) = \prod_{t=0}^{T-1} p_{\theta}(\rvz_{t}| \rvz_{t+1}, \rvy_{0:t})
%     \end{aligned}
% \end{equation*}
where we define $\bar{p}_\theta (\rvy_0 | \rvz_t) \coloneqq p_\theta(\rvy_0 | \bar{\rvz}_0(\rvz_t)) = \normal(\rvy_0 | \gA(\gD(\bar{\rvz}_0(\rvz_t))), (1 - \bar{\alpha}_t)\rmI))$, with the variance taken to be the variance of $\rvz_t | \rvz_0$ according to the forward process. Importantly, as the final target distribution at $t=0$ matches the desired distribution $p_\theta(\rvz_{0:T} | \rvy_{0:T})$ (see \Eqref{eq:target_t0} in Appendix \ref{sec:asymptotic_acc}), in the large compute limit samples from the correct target can be obtained despite all approximations.

\setlength{\tabcolsep}{3pt}
\begin{table*}[!t]
\small
\centering
\caption{Quantitative results on $1024$ examples of size $256 \times 256$ from FFHQ test set. All methods were evaluated under the same experimental setup using LDM. Lower ($\downarrow$) is better in all metrics.}
\begin{tabular}{l ccc c ccc c ccc c ccc}
    \toprule
    & \multicolumn{3}{c}{Inpainting (Box)} && 
    \multicolumn{3}{c}{Inpainting (Free-Form)} &&
    \multicolumn{3}{c}{Gaussian Deblurring} &&
    \multicolumn{3}{c}{Super-Resolution ($8 \times $)}\\
    \cmidrule(l){2-4}  \cmidrule(l){5-8} \cmidrule(l){9-12} \cmidrule(l){13-16}
    & FID & NIQE & LPIPS && FID & NIQE & LPIPS && FID & NIQE & LPIPS && FID & NIQE & LPIPS \\
    \midrule
    LDPS & 39.81 & 7.592 & 0.236 && 40.17 & 7.609 & 0.212 &&  \underline{29.30} & \underline{6.538} & 0.237 && \textbf{29.64} & \textbf{6.412} & 0.282\\
    LTDS & 39.57 & 7.602 & 0.236 && 39.78 & 7.578 & 0.212 && 30.23 & 6.553 & 0.238 && 30.45 & \textbf{6.412} & 0.284\\
    ReSample & 86.79 & 7.142 & 0.230 && 37.01 & \textbf{6.622} & \textbf{0.151} &&  39.80 & 7.441 & 0.275 && 59.23 & 7.307 & 0.356\\
    PSLD & 39.68 & \underline{6.544} & 0.246 && 36.26 & 6.835 & 0.216 && 36.31 & 6.802 & 0.341 && 40.33 &  6.803 & 0.347\\
    PFLD & 39.06 & \textbf{6.509} & 0.245 && 36.43 & \underline{6.817} & 0.215 && 37.16 & 6.751 & 0.343 && 38.11 & 6.832 & 0.345\\ 
    LatentDAPS & 60.24 & 9.999 & 0.257 && 54.40 & 8.766 & 0.223 && 54.28 & 9.496 & 0.283 &&  70.24 & 10.17 & 0.344\\ 
    \midrule
    \MN{} (1 particle) & \textbf{33.37} & 7.032 & \underline{0.212} && \underline{33.67} & 7.034 & \underline{0.194} && \textbf{29.19} & 6.575 & \textbf{0.232} && \underline{30.02} & 6.426 & \textbf{0.277}\\ 
    \MN{} (5 particles) & \underline{33.87} & 7.066 & \textbf{0.211} && \textbf{33.60} & 7.021 & \underline{0.194} && 29.47 & \textbf{6.528} & \underline{0.233} && 30.62 & 6.455 & \underline{0.278}\\ 
    \bottomrule
    \end{tabular}
\label{tab:ffhq}
\end{table*}

\textbf{Sequence weights and proposal distributions.} To derive an SMC procedure using the proposed target distributions, it is common to use importance sampling. The key idea is to construct proposal distributions $\pi_t(\rvz_{t:T})$, one for each timestep, from which it is easy to sample, and approximate $p(\rvz_{t:T} | \rvy_{0:T}) $ with the empirical distribution $\sum_{i=1}^{N} w_t^{(i)} \delta_{\rvz_{t:T}^{(i)}}(\rvz_{t:T})$. Where, $\delta_{\rvz_{t:T}}$ is the Dirac measure and the weight $w_t$ (presented here for $t < T$) is defined as,
\begin{equation}
    \begin{aligned}
        w_t &\propto \frac{\tilde{p}_\theta(\rvz_{t:T} | \rvy_{t:T}, \rvy_0)}{\pi_t(\rvz_{t:T})} \\
        &= \frac{\tilde{p}_\theta(\rvz_{t} | \rvz_{t+1}, \rvy_{t}, \rvy_0)}{\pi_t(\rvz_{t} | \rvz_{t+1})} \frac{\tilde{p}_\theta(\rvz_{t+1:T} | \rvy_{t+1:T}, \rvy_0)}{\pi_{t+1}(\rvz_{t+1:T})}\\
        &= \frac{\tilde{p}_\theta(\rvz_{t} | \rvz_{t+1}, \rvy_{t}, \rvy_0)}{\pi_t(\rvz_{t} | \rvz_{t+1})} w_{t+1}.
    \end{aligned}
    \label{eq:weight_recursion}
\end{equation}
Here, $\tilde{p}_{\theta}(\cdot)$ is the unnormalized approximate target density. Plugging the unnormalized target densities from \Eqref{eq:approx_target} in \Eqref{eq:weight_recursion} for all timesteps $t$ results in,
% \begin{equation}
%     \begin{cases}
%     w_T \propto  \frac{p_\theta(\rvy_T | \rvz_T) \bar{p}_\theta(\rvy_0 | \rvz_T) p(\rvz_{T})}{\pi_{T}(\rvz_{T})}, &t = T, \\
%     w_t \propto \frac{p_\theta(\rvy_t | \rvz_t) \bar{p}_\theta(\rvy_0 | \rvz_t) p_\theta(\rvz_t | \rvz_{t+1})}{\bar{p}_\theta(\rvy_0 | \rvz_{t+1}) \pi_{t}(\rvz_{t} | \rvz_{t+1})}, &0 < t < T,\\
%      w_0 \propto \frac{p_\theta(\rvy_0 |\rvz_0) p_\theta(\rvz_0 | \rvz_{1})}{\bar{p}_\theta(\rvy_0 | \rvz_{1}) \pi_{0}(\rvz_{0} | \rvz_{1})}, &t = 0.
% \end{cases}
% \label{eq:weight_functions}
% \end{equation}
\begin{equation}
%\begin{aligned}   
w_t \propto 
    \begin{cases}
   \frac{p_\theta(\rvy_T | \rvz_T) \bar{p}_\theta(\rvy_0 | \rvz_T) p_\theta(\rvz_{T})}{\pi_{T}(\rvz_{T})}, &t = T, \\
    \frac{p_\theta(\rvy_t | \rvz_t) \bar{p}_\theta(\rvy_0 | \rvz_t) p_\theta(\rvz_t | \rvz_{t+1})}{\bar{p}_\theta(\rvy_0 | \rvz_{t+1}) \pi_{t}(\rvz_{t} | \rvz_{t+1})} w_{t+1}, &0 < t < T,\\
    \frac{p_\theta(\rvy_0 |\rvz_0) p_\theta(\rvz_0 | \rvz_{1})}{\bar{p}_\theta(\rvy_0 | \rvz_{1}) \pi_{0}(\rvz_{0} | \rvz_{1})} w_{1}, &t = 0.
\end{cases}
%\end{aligned}
\label{eq:weight_functions}
\end{equation}

With these weights, we now turn to defining the proposal distributions $\pi_t$ for all timesteps $t$. The optimal choice in the sense of minimizing the variance of the weights is $\pi(\rvz_{t} | \rvz_{t:T}) = p_\theta(\rvz_{t} | \rvz_{t+1}, \rvy_{0:t})$ \cite{murphy2001rao, sarkka2013bayesian}. However, they cannot be obtained in closed form. Hence, we design alternative proposal distributions. The proposal distributions for timesteps $t < T$ are defined to be Gaussian $\pi_t(\rvz_{t} | \rvz_{t+1}) = \normal(\rvm_{t}, \rmS_{t})$ with parameters:
\begin{equation}
\begin{aligned} 
    \rmS_{t} &= \tilde{\sigma}_{t+1}^2 \rmI\\
    \rvm_{t} &= \rvmu_\theta(\rvz_{t+1}, {t+1}) - (\gamma_{t} \nabla_{\rvz_{t+1}} \log~\bar{p}_{\theta}(\rvy_0 | \rvz_{t+1}) \\
    &~~~+ \lambda_{t} \nabla_{\rvmu_\theta(\rvz_{t+1}, {t+1})}\log~q_{\theta}(\rvy_{t} | \rvz_{t+1})).
\end{aligned}
\label{eq:proposal}
\end{equation}
Where $q_{\theta}(\rvy_{t} | \rvz_{t+1}) = \normal(\rvy_{t} | \gA(\gD(\rvmu_\theta(\rvz_{t+1}, {t+1}))), \tau^2 \rmI)$, and $\gamma_t$, $\lambda_t$ are finite scaling coefficients. In addition, $\pi_{T}(\rvz_{T})$, the distribution of the proposal at time $T$, is set to the diffusion prior distribution, $\pi_T(\rvz_T) = \normal(0, \rmI)$.

In \Eqref{eq:proposal}, we set the variance to be the variance of the prior diffusion model, namely $\tilde{\sigma}_{t+1} = \sigma_{t+1}$; however, other choices are also applicable. The idea behind the proposal mean is to correct the prior mean estimation by shifting it toward latents that agree more strongly with both $\rvy_t$ and $\rvy_0$. Specifically, the second term of the correction can be seen as making one gradient update step starting from the current prior mean, which serves as an estimate to $\rvz_t$.

The parameters $\gamma_t$ and $\lambda_t$ control the effect of the correction terms to the prior mean. In practice, it is challenging to control the trade-off between the two correction terms. Hence, we propose the following simple approach. During the sampling process, the first term only is used, that is, $\lambda_t = 0$, and then starting from some predefined timestep $s$, the second term is used as well. The intuition here is that during the initial sampling steps, the quality of the labels $\rvy_t$ may not be good. Therefore, we rely on the first term through the posterior mean estimator to capture the general semantics of the image. However, in later sampling stages the quality of the labels increases (see \Figref{fig:in_forward_process} in the appendix), and the latter correction term can help capture fine details in the image. This design choice also relates to the three-stage phenomenon in the diffusion sampling process witnessed in the literature \cite{yu2023freedom}. Setting $\lambda_t = 0$ for all timesteps reduces the \MN{} proposal update to that of TDS \cite{wu2024practical}. \Appref{sec:scaling_coefs} shows an instantiation of $\gamma_t$ and $\lambda_t$ used in this study.

Having defined the weights and the proposal distributions, we present \MN{} concisely in Algorithm \ref{algo:LD_SMC}. The algorithm has two parts; the first part is the Gibbs sampling process and the second is the SMC sampling using SISR. In the algorithm the abbreviation ''Multi" refers to Multinomial distribution and a resampling step is performed at each iteration, although it is not mandatory. Appendix \ref{sec:asymptotic_acc} provides a proof that LD-SMC can achieve an arbitrarily accurate estimate of $p_\theta(\rvz_0 | \rvy_{0})$ under several conditions. Importantly, the estimate does not depend on any of the approximations made to derive our model. The following Theorem summarizes that informally,

\begin{theorem}
    (informal) Let $\sP_N(\rvz_{0:T}) = \sum_{i=1}^N w^{(i)}_0 \delta_{\rvz_{0:T}^{(i)}}(\rvz_{0:T})$ be the discrete measure obtained by the function $\mathbf{SISR}$ in Algorithm \ref{algo:LD_SMC}. Under regularity conditions $\sP_N(\rvz_{0:T})$ converges setwise to $p_\theta(\rvz_{0:T} | \rvy_{0:T})$ as $N \rightarrow \infty$. Furthermore, the stationary distribution of the Gibbs sampling process is $p_\theta(\rvz_{0:T}, \rvy_{1:T} | \rvy_0)$, and the marginal $p_\theta(\rvz_0 | \rvy_0)$ is the limiting distribution of the $\rvz_0$'s subchain.
\end{theorem}

\textbf{Connection to TDS.} While the derivation is different, the SMC procedure in \MN{} resembles somewhat that of TDS \cite{wu2024practical}. The main difference between the two methods is the dependence on the auxiliary variables $\rvy_{1:T}$, which are only relevant for \MN{} in both the proposal distributions and the weights.  Empirically, we observed that this additional conditioning helped to better align the sampling with the corrupted image $\rvy_0$ compared to using the posterior mean approximation as in \cite{chungdiffusion} and \cite{wu2024practical}. %We stress that other parts of \MN{} are unique only to it, such as the Gibbs procedure, and proposal 

\section{Experiments}

\begin{table*}[!t]
\footnotesize
\centering
\caption{Quantitative results on $1024$ examples of size $256 \times 256$ from ImageNet test set. All methods were evaluated under the same experimental setup using LDM. Lower ($\downarrow$) is better in all metrics.}
\begin{tabular}{l ccc c ccc c ccc c ccc}
    \toprule
    & \multicolumn{3}{c}{Inpainting (Box)} && 
    \multicolumn{3}{c}{Inpainting (Free-Form)} &&
    \multicolumn{3}{c}{Gaussian Deblurring} &&
    \multicolumn{3}{c}{Super-Resolution ($8 \times $)}\\
    \cmidrule(l){2-4}  \cmidrule(l){5-8} \cmidrule(l){9-12} \cmidrule(l){13-16}
    & FID & NIQE & LPIPS && FID & NIQE & LPIPS && FID & NIQE & LPIPS && FID & NIQE & LPIPS \\
    \midrule
    LDPS & 65.04 & 7.935 & 0.379 && 53.47 & 7.867 & 0.334 && 52.48 & 6.855 & 0.383 && 61.02 &  6.514 & 0.439\\
    LTDS & 64.74 & 7.907 & 0.378 && 52.75 & 7.884 & 0.334 && \underline{50.82} & \underline{6.695} & \underline{0.379} && 59.12 & 6.270 & 0.435\\
    ReSample & 90.32 & 8.464 & \textbf{0.318} && 44.15 & 7.104 & \textbf{0.248} && \textbf{46.45} & 7.411 & \textbf{0.353} && 87.65 & 8.290 & 0.491\\
    PSLD & 71.15 & 8.042 & 0.434 && 62.38 & 8.037 & 0.411 && 60.68 & \textbf{6.599} & 0.417 && 66.56 & 7.669 & 0.489\\
    PFLD & 72.83 & 7.933 & 0.436 && 63.34 & 8.026 & 0.414 && 60.94 & 6.733 & 0.421 && 64.72 & 7.685 & 0.492\\ 
    LatentDAPS & 98.24 & 11.36 & 0.394 && 68.65 & 9.625 & 0.330 && 77.09 & 10.55 & 0.417 && 104.6 & 12.68 & 0.489\\ 
    \midrule
    \MN{} (1 particle) & \underline{51.49} & \textbf{6.878} & 0.326 && \underline{38.21} & \underline{6.969} & 0.285 && 52.48 & 6.855 & 0.383 && \underline{58.06} & \underline{6.243} & \underline{0.434}\\ 
    \MN{} (5 particles) & \textbf{50.67} & \underline{6.891} & \underline{0.325} && \textbf{36.18} &  \textbf{6.671} & \underline{0.278} && 52.29 & 6.791 & 0.383 && \textbf{57.89} & \textbf{6.238} & \textbf{0.433}\\ 
    \bottomrule
    \end{tabular}
\label{tab:imagenet}
\end{table*}
\setlength{\tabcolsep}{4pt}

% \begin{table*}[!t]
% %\small
% \caption{Quantitative results on $1024$ examples of size $256 \times 256$ from FFHQ test set. All methods were evaluated under the same experimental setup using LDM.} %Per method, best hyper-parameter configuration was chosen based on the FID.}
% \begin{adjustbox}{width=\linewidth,center}
% %\centering
% \begin{tabular}{l cc c cc c cc c cc}
%     \toprule
%     & \multicolumn{2}{c}{Inpainting (Box)} && \multicolumn{2}{c}{Inpainting (Free-Form)} && \multicolumn{2}{c}{Gaussian Deblurring} &&
%     \multicolumn{2}{c}{Super-Resolution ($8\times$)}\\
%     \cmidrule(l){2-3}  \cmidrule(l){5-6} \cmidrule(l){8-9} \cmidrule(l){11-12}
%     & FID ($\downarrow$) & LPIPS ($\downarrow$) && FID ($\downarrow$) & LPIPS ($\downarrow$) && FID ($\downarrow$) & LPIPS ($\downarrow$) && FID ($\downarrow$) & LPIPS ($\downarrow$)\\
%     \midrule
%     LDPS & 39.81 & 0.236 && 40.17 & 0.212 && \textbf{31.81} & 0.285 &&  \textbf{29.64} & \textbf{0.282}\\
%     LTDS & \underline{39.57} & 0.236 && 39.78 & 0.212 && 33.19 & 0.288 && 30.45 & \underline{0.284}\\
%     ReSample & 86.79 & \underline{0.230} && \underline{37.01} & \textbf{0.151} && 39.80 & \textbf{0.275} && 59.23 & 0.356\\
%     PSLD & 47.51 & 0.312 && 46.94 & 0.290 && 36.31 & 0.341 && 40.33 & 0.347\\
%     \midrule
%     \MN{} (Ours) & \textbf{34.98} & \textbf{0.217} && \textbf{34.65} & \underline{0.200} && \underline{32.23} & \underline{0.279} && \underline{30.07} & \textbf{0.282}\\
%     \bottomrule
%     \end{tabular}
% \label{tab:ffhq}
% \end{adjustbox}
% \end{table*}

\label{experiment}
\subsection{Experimental Setting}
We evaluated \MN{} on ImageNet \cite{ILSVRC15} and FFHQ \cite{KarrasLA19}; both are common in the literature of inverse problems \cite{chungdiffusion, dou2024diffusion}. In ImageNet, samples were conditioned on the class label. The guidance scale was fixed to $1.0$ in all our experiments. Results can be improved by adjusting it \cite{rombach2022high}. %Results can be improved for all methods by adjusting the scale, nevertheless any performance gaps between the methods should remain irrespective of that.
%\id{suggestion - By adjusting the scale, results for all methods can be enhanced; however, any differences in performance among the methods are likely to persist regardless.}\ef{just say it was fixed and not fine-tuned to any methods}
Images were resized to $3 \times 256 \times 256$ and normalized to the range $[0, 1]$. 
We used the latent diffusion models VQ-4 / CIN256-V2 \cite{rombach2022high} as the prior model in FFHQ / ImageNet, respectively, with the DDIM diffusion sampler \cite{SongME21}, according to the data split in \cite{esser2021taming}. We sampled $1024$ random images from the validation set of each dataset which were used to evaluate all methods. 
We followed the protocol of \cite{song2023solving} and added Gaussian noise with zero mean and standard deviation $\tau = 0.01$ to the corrupted images. 
Full experimental details are provided in \Appref{sec:full_exp_details}. 

\textbf{Compared methods.} We compare \MN{} with several recent SoTA baseline methods. All methods were evaluated in a similar experimental setup using the same latent diffusion model to ensure fairness in the comparisons. \textbf{(1) DPS} \cite{chungdiffusion}, which introduces correction to the sampling process of the diffusion through the posterior mean estimator; \textbf{(2) TDS} \cite{wu2024practical}, which uses the twisting technique for approximate sequential Monte Carlo sampling; since DPS and TDS were designed for pixel-space diffusion models, we term the adapted versions of them to the latent space LDPS and LTDS, respectively. \textbf{(3) ReSample} \cite{song2023solving}, which applies an optimization procedure during the sampling process to match the approximate posterior mean to the label followed by a resampling step;
\textbf{(4) PSLD} \cite{rout2024solving}, which introduces a correction term to the DPS step to ``glue" $\rvz_0$; \textbf{(5) PFLD} \cite{nazemi2024particle}, a particle filtering approach for LDM based on PSLD update for the proposal distribution; \textbf{(6) LatentDAPS} which decouples the latents in the sampling trajectory.  In all experiments, we evaluated \MN{} with $N=\{1, 5\}$ particles. Using one particle incurs a computational cost similar to that of LDPS (see Table \ref{tab:compute_cost} in the appendix). LTDS and PFLD were evaluated with $N = 5$ particles. All methods were tested on the following inverse problem tasks: box inpainting, Gaussian deblurring, super-resolution ($8 \times$) and free-form inpainting. We followed the setup in \cite{chungdiffusion} for the first three tasks and the setup in \cite{saharia2022palette} for the last one with a few changes to the default hyperparameters. In inpainting tasks, the masks were randomly sampled for each new sample.

\textbf{Evaluation metrics.}  In the main text we report the following metrics, FID \cite{heusel2017gans}, NIQE \cite{mittal2012making}, and LPIPS \cite{zhang2018unreasonable}. Full results including PSNR and SSIM \cite{wang2004image} are deferred to \Appref{sec:full_results}. FID and NIQE are considered perceptual metrics; lower values in them indicate higher perceptual quality. The other metrics can be considered as distortion metrics, which quantify some discrepancy between the generated images and the ground-truth values. Since perceptual metrics and distortion metrics can be in conflict with each other \cite{blau2018perception}, we put more emphasis on perceptual quality. Hence, for all methods, we performed grid search over hyper-parameters and chose the best configuration according to the FID. 

\begin{figure}[!t]
\centering
    \begin{subfigure}[Effect of $s$]{
    \includegraphics[width=0.45\linewidth]{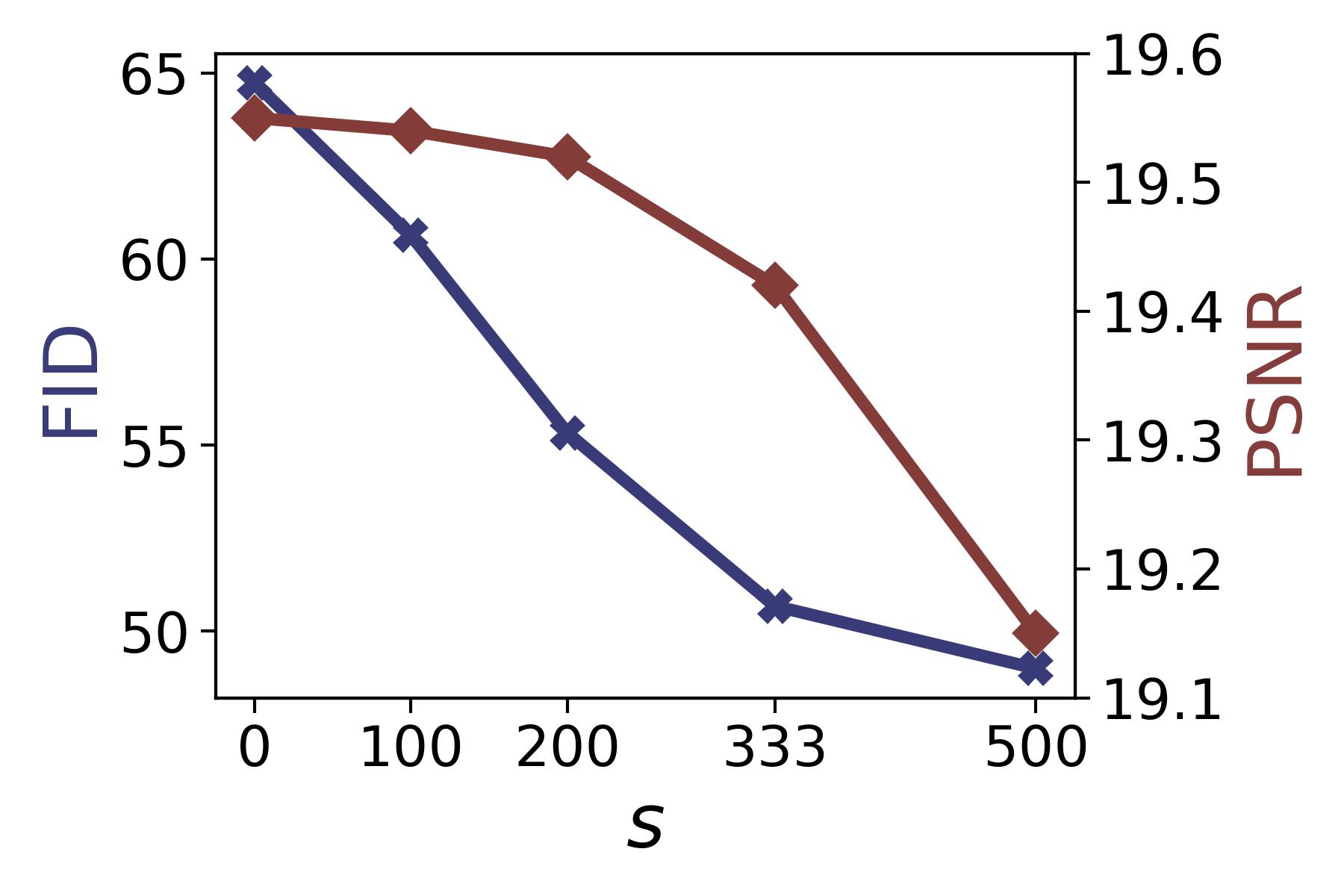}
    \label{fig:ablation_s}
    }
    \end{subfigure}
    %\hfill
    \begin{subfigure}[Effect of $N$]{
    \includegraphics[width=0.45\linewidth]{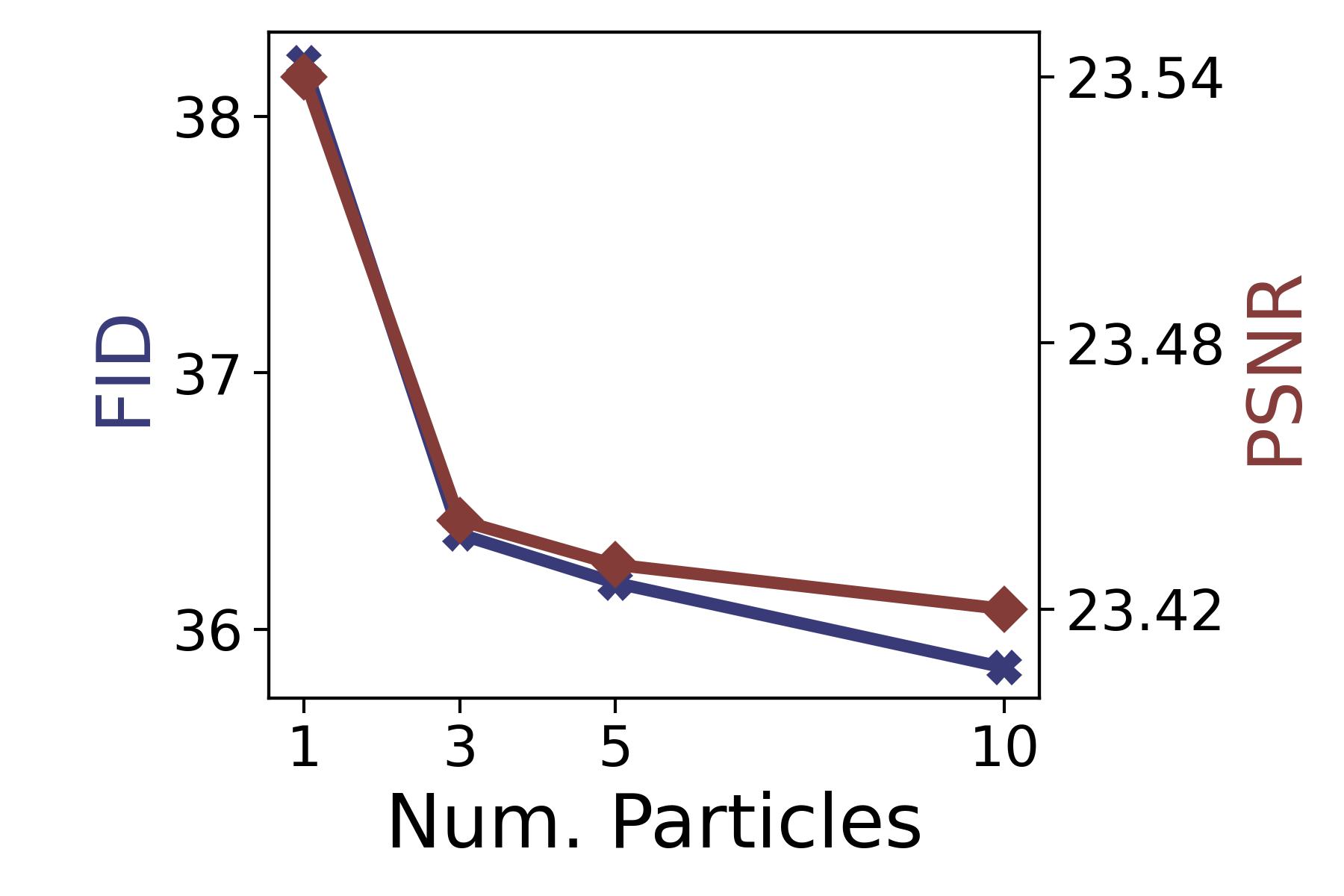}
    \label{fig:ablation_N}
    }
     \end{subfigure}
    \caption{ FID and PSNR values when varying $s$ (left) and the number of particles $N$ (right) on ImageNet box (left) and free-form (right) inpainting tasks.}
    \label{fig:ablation_s_N}
\end{figure}

\subsection{Experimental Results}
Quantitative results are shown in Tables \ref{tab:ffhq} and \ref{tab:imagenet}. From the tables, \MN{} is usually the best or second best among all the comparisons. 
Specifically, on inpainting where extrapolation is needed within the masked region and details should be preserved outside of it, \MN{} can greatly improve over baseline methods. This property is also manifested in \Figref{fig:in_ip_recon} and \Figref{fig:ffhq_ib_recon} in the appendix. \MN{} manages to produce plausible reconstructions while maintaining fine details. The differences are especially highlighted on the ImageNet dataset which has more diversity in it. Also, as is clear from the figures, ReSample images suffer from significant artifacts. We speculate that it partly stems from the complete optimization process performed in every few sampling steps according to this method. % We observe that this method is well suited for some tasks, such as Gaussian deblurring, but does not perform as well on others such as inpainting. 
%was designed specifically for Gaussian deblurring, on which it has the best results, but it doesn't transfer well to image inpainting. % As a result, more emphasis is given to the label and less for the prior diffusion model to generate plausible, natural-looking images. 

\Figref{fig:ffhq_gd_recon_comp} shows qualitative results for Gaussian deblurring of images by all methods. From the figure, \MN{}, LDPS, LTDS, and ReSample are able to generate plausible reconstructions on this task. PSLD and PFLD images have noticeable artifacts, while LatantDAPS images are blurry. 

Additional quantitative and qualitative results can be found in \Appref{sec:full_results} and \ref{sec:img_rec}. In terms of the distortion metrics PSNR and SSIM, \MN{} usually substantially outperforms PSLD and PFLD and has a small advantage over LDPS and LTDS, but overall the three methods are comparable. ReSample and LatentDAPS are usually the best. Nevertheless, as can be seen visually and as indicated by other metrics, it comes at the expense of perceptual quality. LatentDAPS images are overly smooth and details are not preserved in them, while ReSample images have noticeable artifacts, especially in inpainting tasks.

\subsection{Analysis}
%\textbf{Number of particles.}
%\textbf{Proposal update.} 
In \Figref{fig:ablation_s_N} we evaluate \MN{} in terms of FID and PSNR on inpainting tasks when varying $s$, the hyperparameter that controls the influence of $\rvy_t$ in the proposal distribution, and $N$, the number of particles. From \Figref{fig:ablation_s} there is a clear trade-off in FID and PSNR when $s$ changes. Higher $s$ values result in a better FID at the expense of PSNR, while lower $s$ values show the opposite trend. For inpainting, $s=333$ strikes a good balance between the two metrics while giving more emphasis to the perceptual quality. In addition, note that taking $s=0$ is similar to using the DPS as a proposal distribution which results in a significant reduction in perceptual quality. \Figref{fig:ablation_N} shows an improvement in FID while suffering from a small reduction in PSNR when increasing the number of particles. In our comparisons, we chose $N=5$ to balance between performance and computational cost incurred by adding particles (see \Secref{sec:computational_cost} for further details). Empirically, we found that sometimes only increasing the number of particles resulted in favorable metric values, as is also clear from the tables. We attribute this to the relatively small number of particles with which we experimented (up to $10$). However, in general, one can expect that as the number of particles increases, the samples will become more accurate. Finally, in Appendix \ref{sec:further_ablations} we present an ablation study on the effect of using several Gibbs steps. We show that improvements in FID and PSNR can be obtained by taking more than one step, yet there is not always a clear trend of improvements over the steps.

% \Tableref{tab:inpainting_imagenet} presents the results on image inpainting task, and \Tableref{tab:gauss_deblur_imagenet} presents results on Gaussian deblurring task. On the challenging task of inpainting, where extrapolation is needed \MN outperforms all baselines by a large margin in perceptual metrics. On Gaussian deblurring task \MN is second best.

% \subsection{Qualitative Results}
% In \Figref{fig:in_recon} we present qualitative results. As seen from the figure, the completion of \MN is not only sensible but it is able to reconstruct fine details of the image. 

% \section{Limitations}
% Although our approach has strong empirical results,
% one limitation of our approach is related to computational demand. The sampling time and the memory demand increase with the number of particles. In addition, compared to TDS, in the resampling step, we need to use the decoder one more time to compute $p(\rvy_t|\rvz_t)$ (only forward pass), which can also affect the sampling time. This effect can be mitigated by taking fewer particles or by parallelizing \MN{} between GPUs. Furthermore, for more challenging tasks like box inpainting, competing algorithms tend to exhibit noticeable artifacts, which limits their applicability.

\section{Conclusion}
\label{conclusion}
In this study, we presented \MN, a novel method for solving inverse problems in the latent space of diffusion models using SMC. \MN{} augment the model with auxiliary observations, one per each timestep, and use these observations to guide the sampling process as part of the reverse diffusion process. This framework can be seen as applying one step of blocked Gibbs sampling. To perform SMC sampling, we suggested a novel weighing scheme and proposal distributions. Both are based on information from the auxiliary labels and the true label $\rvy_0$.  Empirically, we validated \MN{} against strong baseline methods on common benchmarks. The results suggest that \MN{} can improve the performance over baseline methods, especially in cases where extrapolation is needed (e.g., in inpainting tasks).  one limitation of our approach is related to computational demand. The sampling time and the memory demand increase with the number of particles and Gibbs iterations. This effect can be partially mitigated by taking fewer particles and GPU parallelization.
% \section*{Acknowledgements}

%\newpage
\section*{Impact Statement}
This paper presents work whose goal is to advance the field of Machine Learning. There are many potential societal consequences of our work, none which we feel must be specifically highlighted here.

% In the unusual situation where you want a paper to appear in the
% references without citing it in the main text, use \nocite
% \nocite{langley00}

\bibliography{Ref}
\bibliographystyle{icml2025}

%%%%%%%%%%%%%%%%%%%%%%%%%%%%%%%%%%%%%%%%%%%%%%%%%%%%%%%%%%%%%%%%%%%%%%%%%%%%%%%
%%%%%%%%%%%%%%%%%%%%%%%%%%%%%%%%%%%%%%%%%%%%%%%%%%%%%%%%%%%%%%%%%%%%%%%%%%%%%%%
% APPENDIX
%%%%%%%%%%%%%%%%%%%%%%%%%%%%%%%%%%%%%%%%%%%%%%%%%%%%%%%%%%%%%%%%%%%%%%%%%%%%%%%
%%%%%%%%%%%%%%%%%%%%%%%%%%%%%%%%%%%%%%%%%%%%%%%%%%%%%%%%%%%%%%%%%%%%%%%%%%%%%%%
\newpage
\appendix
% \twocolumn[
% \icmltitle{Appendix of Inverse Problem Sampling in Latent Space \\ Using Sequential Monte Carlo}]
\onecolumn
\icmltitle{Appendix of Inverse Problem Sampling in Latent Space \\ Using Sequential Monte Carlo}

\section{Full Experimental Details}
\label{sec:full_exp_details}
The experiments were carried out mainly using an NVIDIA A100 having 40GB and 80GB memory. 
In all experiments, we used the DDIM formulation \cite{SongME21}, although \MN{} can be applied with other sampling procedures. For all methods, we performed a hyperparameter search on $\eta \in \{0.05, 0.5, 1.0\}$ and found that \MN{} worked best with $\eta = 1.0$. For DPS and TDS we examined several scaling coefficient schemes for the prior mean update, including the ones proposed in each of the corresponding papers, and found that our proposed update worked better for both. For all three methods (\MN{}, DPS, and TDS) we searched for $\kappa_1 \in \{0.3, 0.4, 0.5, 1.0, 1.5, 2.0\}$. For our method, we also performed a grid search over $\kappa_2 \in \{0.5, 1.5, 2.5\}$, $s \in \{0, 100, 200, 333\}$, and $\rho \in \{0.5, 0.75\}$. In addition, when applying resampling, we set the variance of $p_\theta(\rvy_t | \rvz_t)$ to $(1 - \bar{\alpha}_t)\rmI$ to match the variance of $\bar{p}_\theta (\rvy_0 | \rvz_t)$.
For PSLD, in most cases, the default hyperparameters suggested in the paper and code did not yield good results. Hence,
we performed a grid search over PSLD's hyperparameters $\gamma_t \in \{1e-4, 1e-3, 1e-2, 5e-2, 0.1, 0.2\}$ and $\eta_t \in \{0.05, 0.1, 0.2, 0.5, 0.9\}$, and used the best ones for both PSLD and PFLD. For ReSample, we found that using $\eta = 0.0$, the default value in the code, usually performed best. Also, we performed a grid search over $\gamma$, the scaling coefficient of the resampling step std in $\{4, 8, 16, 40, 80, 200, 400\}$. For LatentDAPS, we used the default hyperparameters suggested in the paper, except for super-resolution tasks where we made a grid search over $\eta_0 \in \{1e-4, 5e-5, 1e-5, 5e-6, 1e-6, 5e-7, 1e-7\}$. In addition, for comparability with other methods, we used observation noise of $0.01$, instead of the default $0.05$ in the code, and in ImageNet experiments we used $N=100$ steps.
To choose the best set of hyperparameters, we evaluated each method both visually and using the FID on a sample of $64$ images. Then, we sampled $1024$ images using the best configuration. Similarly to ReSample, we found that applying an optimization process at the end of the sampling process in the latent space can sometimes improve visibility and metric values. We evaluated all models except LatentDAPS with and without this final optimization process and chose the best according to the FID. The optimization procedure was not applied to inpainting tasks since it created non-smooth changes at the boundaries of the box, making the images look non-natural.

For \MN{}, we present here the chosen hyperparameters:
\begin{table*}[!h]
%\small
\centering
\caption{\MN{} hyperparameters for all tasks.}
\begin{tabular}{l l ccccc}
    \toprule
    Dataset & Task & $\kappa_1$ & $\kappa_2$ & $s$ & $\rho$ & Final Latent Optimization\\
    \midrule
    \multirow{ 4}{*}{FFHQ} & Inpainting (Box) & 1.0 & 2.5 & 333 & 0.75 & No\\
    & Inpainting (Free-Form) & 1.0 & 2.5 & 333 & 0.75 & No\\
    & Gaussian deblurring & 1.5 & 1.5 & 100 & 0.5 & Yes\\
    & Super-resolution & 1.0 & 2.5 & 100 & 0.5 & Yes \\
    \midrule
    \multirow{ 4}{*}{ImageNet} & Inpainting (Box) & 2.0 & 2.5 & 333 & 0.75 & No\\
    & Inpainting (Free-Form) & 2.0 & 2.5 & 333 & 0.75 & No\\
    & Gaussian deblurring & 0.5 & -- & 0  & -- & Yes\\
    & Super-resolution & 0.3 & 2.5 & 333 & 0.5 & No\\
    \bottomrule
    \end{tabular}
\label{tab:deafult_hps}
\end{table*}

\section{Approximate Target Distributions}
\label{sec:approx_target_dist}
In \Secref{sec:post_sampling} we presented approximate target distributions for the SMC sampling procedure. Here we present the derivation for each case, namely (1) $t=T$, (2) $0 < t < T$, and (3) $t=0$:

\begin{enumerate}[(1)]
    \item $\underline{t=T:}$
    \begin{equation*}
        \begin{aligned} 
            p_\theta(\rvz_{T} | \rvy_{T}, \rvy_0)
            &\propto p_\theta(\rvy_T | \rvz_{T}, \rvy_0) p_\theta(\rvz_{T} | \rvy_0) \\
            &\propto p_\theta(\rvy_T | \rvz_{T}) p_\theta(\rvy_0 | \rvz_{T})p(\rvz_{T})\\
            &\approx p_\theta(\rvy_T | \rvz_{T}) \bar{p}_\theta(\rvy_0 | \rvz_{T})p(\rvz_{T}).
        \end{aligned}
        \end{equation*}
    Where, in the first transition we used Bayes rule, in the second transition we used the model dependencies and Bayes rule again, and in the last transition we make the additional approximation of conditioning on the posterior mean estimator.
    \item $\underline{0 < t < T:}$
    \begin{equation*}
        \begin{aligned} 
            p_\theta(\rvz_{t:T} | \rvy_{t:T}, \rvy_0)
            &\propto p_\theta(\rvy_t | \rvz_{t:T}, \rvy_{t+1:T}, \rvy_0) p_\theta(\rvz_{t:T} | \rvy_{t+1:T}, \rvy_0) \\
            &= p_\theta(\rvy_t | \rvz_t) p_\theta(\rvz_t | \rvz_{t+1:T}, \rvy_{t+1:T}, \rvy_0)p_\theta(\rvz_{t+1:T} | \rvy_{t+1:T}, \rvy_0) \\
            &= p_\theta(\rvy_t | \rvz_t) p_\theta(\rvz_t | \rvz_{t+1}, \rvy_0) p_\theta(\rvz_{t+1:T} | \rvy_{t+1:T}, \rvy_0) \\
            &= p_\theta(\rvy_t | \rvz_t) \frac{p_\theta(\rvy_0 | \rvz_t)}{p_\theta(\rvy_0 | \rvz_{t+1})}p_\theta(\rvz_t|\rvz_{t+1})p_\theta(\rvz_{t+1:T} | \rvy_{t+1:T}, \rvy_0) \\
            & \approx \frac{p_\theta(\rvy_t | \rvz_t) \bar{p}_\theta(\rvy_0 | \rvz_t)}{\bar{p}_\theta(\rvy_0 | \rvz_{t+1})}p_\theta(\rvz_t|\rvz_{t+1})p_\theta(\rvz_{t+1:T} | \rvy_{t+1:T}, \rvy_0).
        \end{aligned}
        \end{equation*}
        Where, in the first transition we used Bayes rule, in the third transition we used the Markovian assumption, in the forth transition we used Bayes rule again, and in the last transition we make an additional approximation and condition on the posterior mean estimator for both time $t$ and time $t+1$.
    \item $\underline{t=0:}$
    \begin{equation*}
        \begin{aligned} 
            p_\theta(\rvz_{0:T} | \rvy_{1:T}, \rvy_0)
            % &\propto p_\theta(\rvy_0 | \rvz_{0:T}, \rvy_{1:T}) p_\theta(\rvz_{0:T} | \rvy_{1:T}) \\
            % &= p_\theta(\rvy_0 | \rvz_0) p_\theta(\rvz_0 | \rvz_{1:T}, \rvy_{1:T})p_\theta(\rvz_{1:T} | \rvy_{1:T}) \\
            % &= p_\theta(\rvy_0 | \rvz_0) p_\theta(\rvz_0 | \rvz_{1}) p_\theta(\rvz_{1:T} | \rvy_{1:T}).
            &= p_\theta(\rvz_0 | \rvz_{1:T}, \rvy_{0:T}) p_\theta(\rvz_{1:T} | \rvy_{1:T}, \rvy_0) \\
            &= p_\theta(\rvz_0 | \rvz_1, \rvy_0) p_\theta(\rvz_{1:T} | \rvy_{0:T}) \\
            &= \frac{p_\theta(\rvy_0 |\rvz_0)}{p_\theta(\rvy_0 |\rvz_1)}p_\theta(\rvz_0 |\rvz_1) p_\theta(\rvz_{1:T} | \rvy_{0:T}) \\
            & \approx \frac{p_\theta(\rvy_0 |\rvz_0)}{\bar{p}_\theta(\rvy_0 |\rvz_1)}p_\theta(\rvz_0 |\rvz_1) p_\theta(\rvz_{1:T} | \rvy_{0:T})
            % &= p_\theta(\rvy_0 | \rvz_0) p_\theta(\rvz_0 | \rvz_{1}) p_\theta(\rvz_{1:T} | \rvy_{1:T}).
        \end{aligned}
        \end{equation*}
        Where, in the third transition we used Bayes rule, and in the last transition we make an additional approximation and condition on the posterior mean estimator.
\end{enumerate}

\section{Asymptotic Accuracy of \MN}
\label{sec:asymptotic_acc}
Here we provide a proof that $\rvz_0$ samples can be made arbitrarily accurate according to our model design.  Our proof is composed of three parts, all of which have been previously established in the literature. We will restate them here and accommodate them to our setting. The three parts are, (1) augmenting the model with auxiliary random variables, ergo performing a completion of the desired marginal density $p_\theta(\rvz_0 | \rvy_0)$ (Definition 10.3 in \cite{robert1999monte}), (2) Proving asymptotic accuracy of the SMC procedure following a similar lines of Theorem 2 in \cite{wu2024practical}, (3) Showing sufficient conditions that the Markov chain generated by the Gibbs sampling procedure is ergodic and hence $p_\theta(\rvz_0 | \rvy_0)$ is the limiting distribution from which $\rvz_0$'s are sampled (Theorem 10.6 in \cite{robert1999monte}).

First of all, here we provide a concise version of our Gibbs sampler presented in \Secref{sec:sampling_procedure}. We denote by $j$ the index of the Gibbs iterations:
\begin{enumerate}
    \item Initialize $\rvz_{0:T}^{0}$
    \item For $j = 0, ..., J-1$:
    \begin{enumerate}
        \item Sample, $\rvy_{1:T}^{j+1} \sim p_{\theta}(\rvy_{1:T}| \rvz_{0:T}^{j}, \rvy_0)$.
        \item Sample, $\rvz_{0:T}^{j+1} \sim p_{\theta}(\rvz_{0:T}|\rvy_{1:T}^{j+1}, \rvy_0)$ using SMC.
    \end{enumerate}
\end{enumerate}
%sequence of samples generated by our proposed Gibbs sampler. Similarly $(\rvz_0^(j))$ denote the corresponding subchain 

\definition \cite{robert1999monte}. Given a probability density $p_\theta(\rvz_0 | \rvy_0)$, a density $g$
that satisfies $\int g(\rvz_0, \rvz_{1:T}, \rvy_{1:T} | \rvy_0) d\rvz_{1:T} d\rvy_{1:T}$ is called a completion of $p_\theta(\rvz_0 | \rvy_0)$.

We denote by $(\rvz_{0:T}^{j}, \rvy_{1:T}^{j})$ the Markov chain generated by our proposed Gibbs sampler. Similarly, denote by $(\rvz_0^{j})$ the corresponding subchain. 

\begin{theorem}\cite{robert1999monte}.
\label{thm:gibbs}
For the Gibbs sampler described in \Secref{sec:sampling_procedure}, if $(\rvz_{0:T}^{j}, \rvy_{1:T}^{j})$ is ergodic, then the distribution $g$ is a stationary distribution for it and $p_\theta(\rvz_0 | \rvy_0)$ is the limiting distribution of the subchain $(\rvz_0^{j})$. 
\end{theorem}

\textit{Proof.} The support of $g$ is $\sR^d$, with $d = dim(\rvz_{0:T}, \rvy_{1:T})$, and hence is connected. Also, since each conditional distribution in the Gibbs sampling process
is a Gaussian or a multiplication of Gaussian densities, they are all strictly positive. Following Lemma 10.11 in \cite{robert1999monte}, $(\rvz_{0:T}^{j}, \rvy_{1:T}^{j})$ is irreducible and aperiodic, i.e., it is ergodic. \qed

Specifically, Theorem \ref{thm:gibbs} shows that in order to obtain samples from the marginal distribution $p_\theta(\rvz_0 | \rvy_0)$, we need to sample $\rvy_{1:T}$ and $\rvz_{0:T}$ from the conditional distribution according to the model at each iteration. Sampling $\rvy_{1:T}$ variables is straightforward as it requires sampling from independent Gaussian distributions, that is, $p_\theta(\rvy_t | \rvz_t) = \normal(\rvy_t | \gA(\gD(\rvz_t)), \tau^2 \rmI)$ for all $t > 0$. On the other hand, sampling $\rvz_{0:T}$ requires sampling using an SMC procedure. As we show next in the large particle limit, samples from $p_{\theta}(\rvz_{0:T} | \rvy_{0:T})$ are accurate. 

% We note here that due to the posterior approximation of the SMC procedure, $\rvz_{2:T}$ and $\rvy_{1:T}$ samples are not accurate, yet since $\rvz_0$ samples are in fact from the correct posterior (as we will show next), and the subchain $(\rvz_0^{j})$ converges to the marginal distribution $p_\theta(\rvz_0 | \rvy_0)$. 

To prove the asymptotic accuracy of the SMC procedure in \MN, we adopt the formulation of \cite{wu2024practical}. In what follows, to prevent cluttered notation, we drop the index of the Gibbs iteration and note that all quantities are those of a specific iteration. Specifically, here we first reiterate the following $3$ important quantities, the prior distribution $p_\theta(\rvz_{0:T})$, the proposal distributions $\pi_{t}(\rvz_{t} | \rvz_{t+1})$, and the weighting functions $w_t(\rvz_t, \rvz_{t+1})$ and then proceed to the proof. 

\textbf{Prior distribution}. Let $p_\theta(\rvz_{0:T})$ denote the diffusion generative model defined according to \Eqref{eq:ddim_sampling}. Then, the Markovian structure of the prior diffusion model goes as follows,
\[
\begin{cases}
    p_\theta(\rvz_{t} | \rvz_{t+1}) = \normal(\rvmu_\theta(\rvz_{t+1}, {t+1}), \sigma_{t+1}^2 \rmI) \\
    \qquad \qquad \quad = \normal(\rvz_{t} | \sqrt{\bar{\alpha}_{t}}\left(\frac{\rvz_{t+1} - \sqrt{1 - \bar{\alpha}_{t+1}} \cdot \rvepsilon_\theta (\rvz_{t+1}, {t+1})}{\sqrt{\bar{\alpha}_{t+1}}}\right) + \sqrt{1 - \bar{\alpha}_{t} - \sigma^2_{t+1}} \cdot \rvepsilon_\theta (\rvz_{t+1}, {t+1}), \sigma_{t+1}^2 \rmI), &1 < t < T,\\
    %p_\theta(\rvz_{T}) = \normal(\rvz_{T} | \rvz_T, \xi^2 \rmI),   
    p_\theta(\rvz_{T}) = \normal(0, \rmI) & t = T.
\end{cases}
\]
%where $\rvz_{T} \sim \normal(0, \rmI)$.
% $p_\theta(\rvz_{t-1} | \rvz_{t})$ for $t < T$, and $p_\theta(\rvz_{T}$

\textbf{Proposal distributions}. Denote the proposal distribution for timestep $t < T$ as $\pi_{t}(\rvz_{t} | \rvz_{t+1}) = \normal(\rvm_{t}, \rmS_{t})$, with parameters
\begin{equation}
\begin{aligned}
    % \rmS_{t-1} &= \sigma_t^2 \rmI\\
    % \rvm_{t-1} &= \rvmu_\theta(\rvz_t, t) - (\gamma_{t-1} \nabla_{\rvz_t}||\rvy_0 - \gA(\gD(\bar{\rvz}_0(\rvz_t)))||_2^2 + \lambda_{t-1} \nabla_{\rvmu_\theta(\rvz_t, t)}||\rvy_{t-1} - \gA(\gD(\rvmu_\theta(\rvz_t, t)))||_2^2).
    \rmS_{t} &= \tilde{\sigma}_{t+1}^2 \rmI\\
    \rvm_{t} &= \rvmu_\theta(\rvz_{t+1}, {t+1}) - (\gamma_{t} \nabla_{\rvz_{t+1}} \log~\bar{p}_{\theta}(\rvy_0 | \rvz_{t+1}) + \lambda_{t} \nabla_{\rvmu_\theta(\rvz_{t+1}, {t+1})}\log~q_{\theta}(\rvy_{t} | \rvz_{t+1})).
    % \pi_t(\rvz_{t-1} | \rvz_{t}) = \rvmu_\theta(\rvz_t, t) - (\gamma_t \nabla_{\rvz_t} \log~p(\rvy_0 | \bar{\rvz}_0(\rvz_t))
    % + \lambda_t \nabla_{\rvmu_\theta(\rvz_t, t)}||\rvy_t - \gA(\gD(\rvmu_\theta(\rvz_t, t)))||_2^2) 
\end{aligned}
\end{equation}
Where $\bar{p}_{\theta}(\rvy_0 | \rvz_{t+1}) = \normal(\rvy_0 | \gA(\gD(\bar{\rvz}_0(\rvz_{t+1}))), (1 - \bar{\alpha}_{t+1})\rmI)$, and $q_{\theta}(\rvy_{t} | \rvz_{t+1}) = \normal(\rvy_{t} | \gA(\gD(\rvmu_\theta(\rvz_{t+1}, {t+1}))), \tau^2 \rmI)$, and $\gamma_t$, $\lambda_t$ are finite scaling coefficients. The distribution of the proposal at time $T$, $\pi_{T}(\rvz_{T})$, from which the initial sample is taken, is set to the diffusion prior distribution $p_\theta(\rvz_{T}) = \normal(0, \rmI)$.

\textbf{Weighting functions}. The unnormalized weighting functions for all timesteps $t$ are summarized as follows,
\[
\begin{cases}
    \tilde{w}_T(\rvz_T) = p_\theta(\rvy_T | \rvz_T) \bar{p}_\theta(\rvy_0 | \rvz_T), \\
    \tilde{w}_t(\rvz_t, \rvz_{t+1}) = p_\theta(\rvy_t | \rvz_t) \bar{p}_\theta(\rvy_0 | \rvz_t) p_\theta(\rvz_t | \rvz_{t+1}) / (\bar{p}_\theta(\rvy_0 | \rvz_{t+1}) \pi_{t}(\rvz_{t} | \rvz_{t+1})), \\
     \tilde{w}_0(\rvz_0, \rvz_1) = p_\theta(\rvy_0 |\rvz_0) p_\theta(\rvz_0 | \rvz_{1}) / \bar{p}_\theta(\rvy_0 | \rvz_{1}) \pi_{0}(\rvz_{0} | \rvz_{1}),
\end{cases}
\]
where $p_\theta(\rvy_t | \rvz_t) = \normal(\rvy_t | \gA(\gD(\rvz_t)), \tau^2\rmI)$.

In SMC the proposal distributions and weighting functions define a sequence of intermediate target distributions $\{\nu_t\}_{t=0}^T$ defined as follows,

\begin{equation}
\begin{aligned}
   \nu_t(\rvz_{t:T}) &= \frac{1}{\gL_t}\left( \pi_T(\rvz_T) \prod_{t'=t}^{T} \pi_{t'}(\rvz_{t'} | \rvz_{t'+1}) \right) \left(\tilde{w}_T(\rvz_T) \prod_{t'=t}^{T} \tilde{w}_{t'}(\rvz_{t'}, \rvz_{t'+1})\right).
\end{aligned}
\label{eq:smc_targets}
\end{equation}

Importantly, if for given weighting functions and proposal distributions the final target distribution $\nu_0$ coincides with the desired posterior distribution $p(\rvz_{0:T} | \rvy_{0:T})$, then samples that are approximately distributed according to $p(\rvz_{0:T} | \rvy_{0:T})$ can be obtained \cite{doucet2001introduction, naesseth2019elements}. The approximation becomes accurate in the limit of large number of particles. For $t=0$, plugging in \Eqref{eq:smc_targets} the proposal distributions and weighting functions and observing that all appearances of the proposal distributions besides that of time $T$ cancel out, we obtain:
\begin{equation}
\begin{aligned}
   \nu_0(\rvz_{0:T}) &= \frac{1}{\gL_0} p_\theta(\rvz_T) p_\theta(\rvy_T | \rvz_T) \bar{p}_\theta(\rvy_0 | \rvz_T) \left[ \prod_{t=1}^{T-1} \frac{p_\theta(\rvy_t | \rvz_t) \bar{p}_\theta(\rvy_0 | \rvz_t) p_\theta(\rvz_t | \rvz_{t+1})}{\bar{p}_\theta(\rvy_0 | \rvz_{t+1})}\right] \frac{p_\theta(\rvy_0 |\rvz_0)}{\bar{p}_\theta(\rvy_0 |\rvz_1)} p_\theta(\rvz_0 | \rvz_{1})\\
   &= \frac{1}{\gL_0} \prod_{t=0}^{T} p_\theta(\rvy_t | \rvz_t) \prod_{t=0}^{T-1} p_\theta(\rvz_t | \rvz_{t+1}) p_\theta(\rvz_T) = \frac{1}{\gL_0} p_\theta(\rvy_{0:T} | \rvz_{0:T}) p_\theta(\rvz_{0:T}) = p_\theta(\rvz_{0:T} | \rvy_{0:T}).
\end{aligned}
\label{eq:target_t0}
\end{equation}
Where in the second equality $\bar{p}_{\theta}$ terms cancel out and in the last equality we applied Bayes rule.

Now that we established necessary conditions to sample from $p_\theta(\rvz_{0:T} | \rvy_{0:T})$, the following theorem from \cite{chopin2020introduction} characterizes the conditions under which SMC algorithms converge. We adopt the formulation of \cite{wu2024practical} and present it here, adapted to our case.

\begin{theorem}
\label{thm:convergence}
     (\cite{chopin2020introduction} – Proposition 11.4). Let $\{\rvz_{0:T}^{(i)}, w_0^{(i)}\}$ be the sequence of particles and final weights returned by the last iteration of the SMC algorithm with $N$ particles and using multinomial resampling. If the weight functions of all timesteps $w_t^{(i)}$ are positive and bounded, then for every $\nu_0$-measurable function $\phi$ of $\rvz_{0:T}^{(i)}$,
     \begin{equation*}
        \lim_{N \rightarrow \infty} \sum_{i=1}^{N} w_0^{(i)} \phi(\rvz_{0:T}^{(i)}) = \int \phi(\rvz_{0:T}) \nu_0(\rvz_{0:T}) d\rvz_{0:T}
     \end{equation*}
     with probability one.
\end{theorem}

Specifically, taking $\phi(\rvz_{0:T}) = \sI[\rvz_{0:T} \in E]$ for any $\nu_0$-measurable set E implies the convergence of $\sP_N(E) = \sum_{i=1}^N w^{(i)}_0 \delta_{\rvz_{0:T}^{(i)}}(E)$. Here, $\delta_\rvz$ is the Dirac measure defined for a given point $\rvz$ and a $\nu_0$-measurable set E. The following proposition characterizes the conditions under which Theorem \ref{thm:convergence} applies in our case.

\begin{proposition}
Let $\sP_N(E) = \sum_{i=1}^N w^{(i)}_0 \delta_{\rvz_{0:T}^{(i)}}(E)$ be a discrete distribution over particles with $\{(\rvz_{0:T}^{(i)}, w^{(i)}_0)\}_{i=1}^N$ returned by the SMC procedure in \Secref{sec:post_sampling} with $N$ particles. Assume for all $t$:
\begin{enumerate}[(a)]
    \item The likelihood functions $p_{\theta}(\rvy_t | \rvz_{t}) = \normal(\rvy_t | \gA(\gD(\rvz_t)), \tau^2\rmI)$, the ratios $\bar{p}_{\theta}(\rvy_0 | \rvz_{t})/\bar{p}_{\theta}(\rvy_0 | \rvz_{t+1})$, and $\bar{p}_{\theta}(\rvy_0 | \rvz_{T})$, are all positive and bounded.
    \item $\log \bar{p}_{\theta}(\rvy_0 | \rvz_{t})$ is continuous and has bounded gradients in $\rvz_{t}$.
    \item $\log q_{\theta}(\rvy_{t} | \rvz_{t+1})$ is continuous and has bounded gradients in $\rvmu_\theta(\rvz_{t+1}, t+1)$.
    \item The proposal variance is larger than the prior diffusion model variance, namely  $\tilde{\sigma}_t^2 > \sigma_t^2$.
\end{enumerate}
Then $\sP_N$ converges setwise to $p_{\theta}(\rvz_{0:T}|\rvy_{0:T})$ with probability one, that is for every set $E$, $\lim_{N \rightarrow \infty}\sP_N(E) = \int_{E} p_{\theta}(\rvz_{0:T}|\rvy_{0:T}) d\rvz_{0:T}$.
\end{proposition}
Note that here, unlike \cite{wu2024practical}, we do not need to assume that $\bar{p}_\theta(\rvy_0 | \rvz_0) = p_{\theta}(\rvy_0 | \rvz_0)$ due to our revised posterior distribution. Here, however, we add assumption (c). Assumptions (b) and (c) are the strongest assumptions as they may not apply even for linear transformations. But, for sufficiently smooth decoder and diffusion model in the input ($\rvz_t$ or $\rvmu_\theta(\rvz_{t+1}, t+1)$) with uniformly bounded gradients, the assumptions will hold. 

\textit{Proof.}
%Similar to \cite{wu2024practical} we rely on Proposition 11.4 in \cite{chopin2020introduction} which presents sufficient conditions for the setwise convergence of $\sP_N = \sum_{i=1}^N w^{(i)}_0 \delta_{\rvz_0^{(i)}}$. Specifically, we need to show that (1) the target distribution of the SMC procedure at time $t=0$ is $p_\theta(\rvz_{0:T} | \rvy_{0:T})$, and (2) the weighting functions $w_t \coloneqq w_t(\rvz_t, \rvz_{t+1})$ of all timesteps are all positive and bounded.
To prove the statement, we need to show that (1) the target distribution of the SMC procedure at time $t=0$ is $p_\theta(\rvz_{0:T} | \rvy_{0:T})$, and (2) the weighting functions $w_t(\rvz_t, \rvz_{t+1})$ of all timesteps are all positive and bounded. The first condition was already established in \Eqref{eq:target_t0}, hence we proceed to the second condition.

Since all the weights are defined through multiplications of density functions, they are strictly positive. To show that they are bounded, we first note that, by assumption (a), $w_T(\rvz_T)$ is bounded. To show that the intermediate weights are bounded it is enough to consider the log transformation of the unnormalized weights,
\[
\log~\tilde{w}_t(\rvz_t, \rvz_{t+1}) = 
\begin{cases}
    \log~p_\theta(\rvy_t | \rvz_t) + \log~\frac{\bar{p}_\theta(\rvy_0 | \rvz_t)}{\bar{p}_\theta(\rvy_0 | \rvz_{t+1})} + \log~\frac{p_\theta(\rvz_t | \rvz_{t+1})}{\pi_{t}(\rvz_{t} | \rvz_{t+1})}, &0 < t < T,\\
     \log~p_\theta(\rvy_0 |\rvz_0) + \log~\frac{p_\theta(\rvz_0 | \rvz_{1})}{\pi_{0}(\rvz_{0} | \rvz_{1})}, & t = 0.
\end{cases}
\]

By assumption (a), $\log~p_\theta(\rvy_t | \rvz_t)$ and $\log~\bar{p}_\theta(\rvy_0 | \rvz_t) / \bar{p}_\theta(\rvy_0 | \rvz_{t+1})$ are bounded for all $t$. We will show next that the $\log~p_\theta(\rvz_t | \rvz_{t+1}) / \pi_{t}(\rvz_{t} | \rvz_{t+1})$ terms are all bounded based on assumptions (b) - (d). We denote by $\rvmu_t \coloneqq \rvmu_\theta(\rvz_{t+1}, {t+1})$,

\begin{equation*}
    \begin{aligned}
        \log~\frac{p_\theta(\rvz_t | \rvz_{t+1})}{\pi_{t}(\rvz_{t} | \rvz_{t+1})} &= \log~\frac{|2\pi\sigma_{t+1}^2\rmI|^{-0.5} \exp\{-(2 \sigma_{t+1}^2)^{-1}||\rvz_t - \rvmu_t||_2^2\}}{|2\pi\tilde{\sigma}_{t+1}^2\rmI|^{-0.5} \exp\{-(2 \tilde{\sigma}_{t+1}^2)^{-1}||\rvz_t - \rvm_t||_2^2\}}\\
        &\consteq -\frac{1}{2}[ \sigma_{t+1}^{-2}||\rvz_t - \rvmu_t||_2^2 - \tilde{\sigma}_{t+1}^{-2}||\rvz_t - \rvm_t||_2^2]\\
        &\text{Expanding, } ||\rvz_t - \rvm_t \pm \rvmu_t||_2^2 = ||\rvz_t - \rvmu_t||^2_2 + 2\langle \rvz_t - \rvmu_t,\rvmu_t - \rvm_t \rangle + ||\rvmu_t - \rvm_t||_2^2\\
        &= -\frac{1}{2}[ (\sigma_{t+1}^{-2} - \tilde{\sigma}_{t+1}^{-2})||\rvz_t - \rvmu_t||_2^2 - 2\tilde{\sigma}_{t+1}^{-2} \langle \rvz_t - \rvmu_t,\rvmu_t - \rvm_t \rangle - \tilde{\sigma}_{t+1}^{-2} ||\rvmu_t - \rvm_t||_2^2]\\
        & \text{Notice, } ||\rvmu_t - \rvm_t||_2^2 = ||\gamma_{t} \nabla_{\rvz_{t+1}} \log~\bar{p}_{\theta}(\rvy_0 | \rvz_{t+1}) + \lambda_{t} \nabla_{\rvmu_t}\log~q_{\theta}(\rvy_{t} | \rvz_{t+1})||_2^2 < \infty\\ &\text{ by applying Minkowski inequality and using assumptions (b) and (c)}\\
        &\consteq  -\frac{1}{2}[ (\sigma_{t+1}^{-2} - \tilde{\sigma}_{t+1}^{-2})||\rvz_t - \rvmu_t||_2^2 - 2\tilde{\sigma}_{t+1}^{-2} \langle \rvz_t - \rvmu_t,\rvmu_t - \rvm_t \rangle]\\
        &\text{Apply Cauchy-Schwartz inequality,} \\
        &\leq -\frac{1}{2} (\sigma_{t+1}^{-2} - \tilde{\sigma}_{t+1}^{-2})||\rvz_t - \rvmu_t||_2^2 + \tilde{\sigma}_{t+1}^{-2} ||\rvmu_t - \rvm_t||_2 ||\rvz_t - \rvmu_t||_2 ]\\
        & \text{Apply the inequality } -\frac{a}{2} x^2 + bx \leq \frac{b^2}{2a} \text{for } a = \sigma_{t+1}^{-2} - \tilde{\sigma}_{t+1}^{-2} > 0 \text{ by assumption (d)}\\
        &\leq \frac{\tilde{\sigma}_{t+1}^{-4}||\rvmu_t - \rvm_t||_2^2}{2(\sigma_{t+1}^{-2} - \tilde{\sigma}_{t+1}^{-2})} < \infty.
    \end{aligned}
\end{equation*}
Where the last inequality is again due to assumptions (b) and (c). Here, $\consteq$ denotes an equality up to a constant. \qed

\begin{figure}[!t]
\centering
    \begin{subfigure}[Super-Resolution on ImageNet]{
    \includegraphics[width=0.4\linewidth]{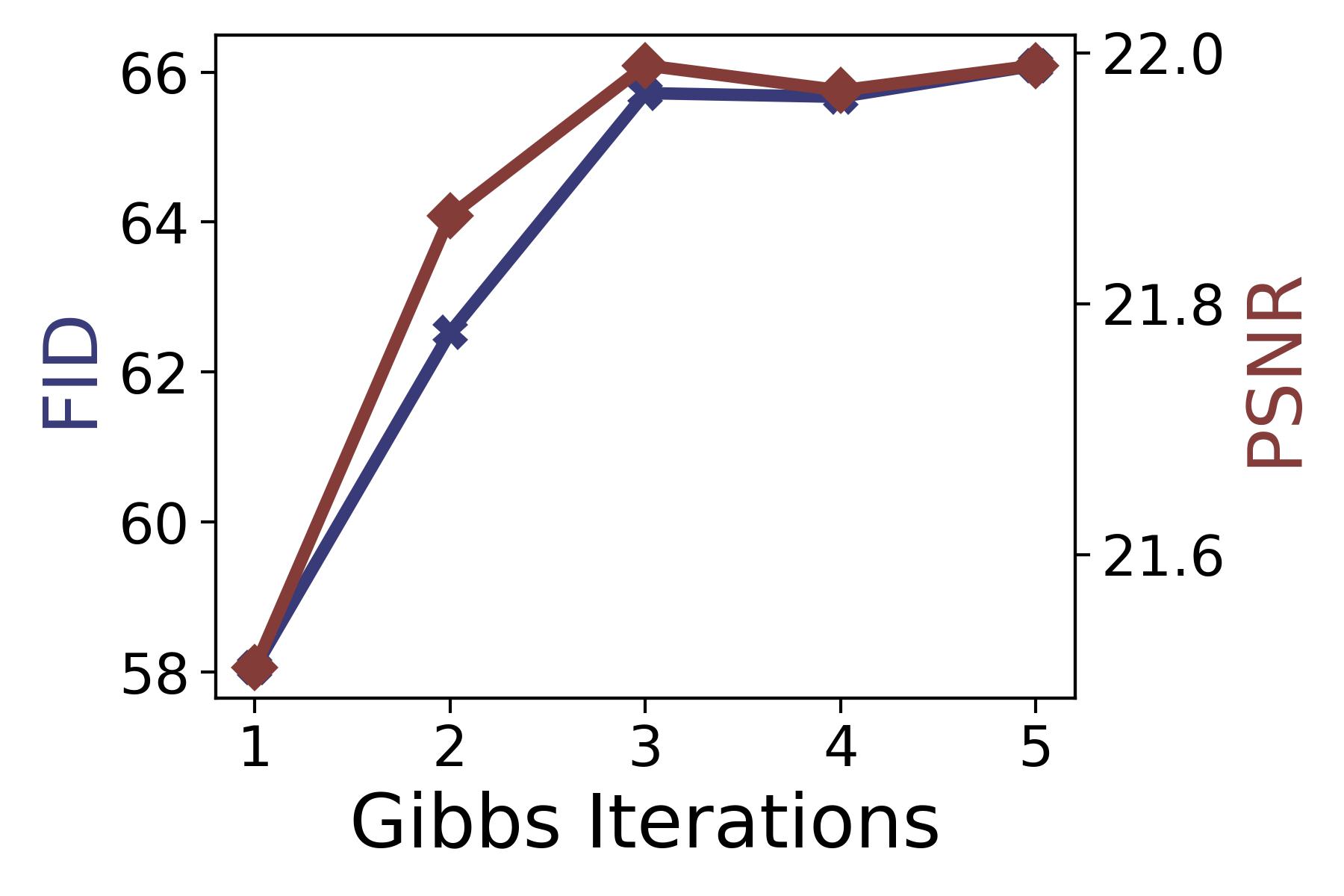}
    \label{fig:gibbs_sr}
    }
    \end{subfigure}
    %\hfill
    \begin{subfigure}[Gaussian Deblurring on ImageNet]{
    \includegraphics[width=0.4\linewidth]{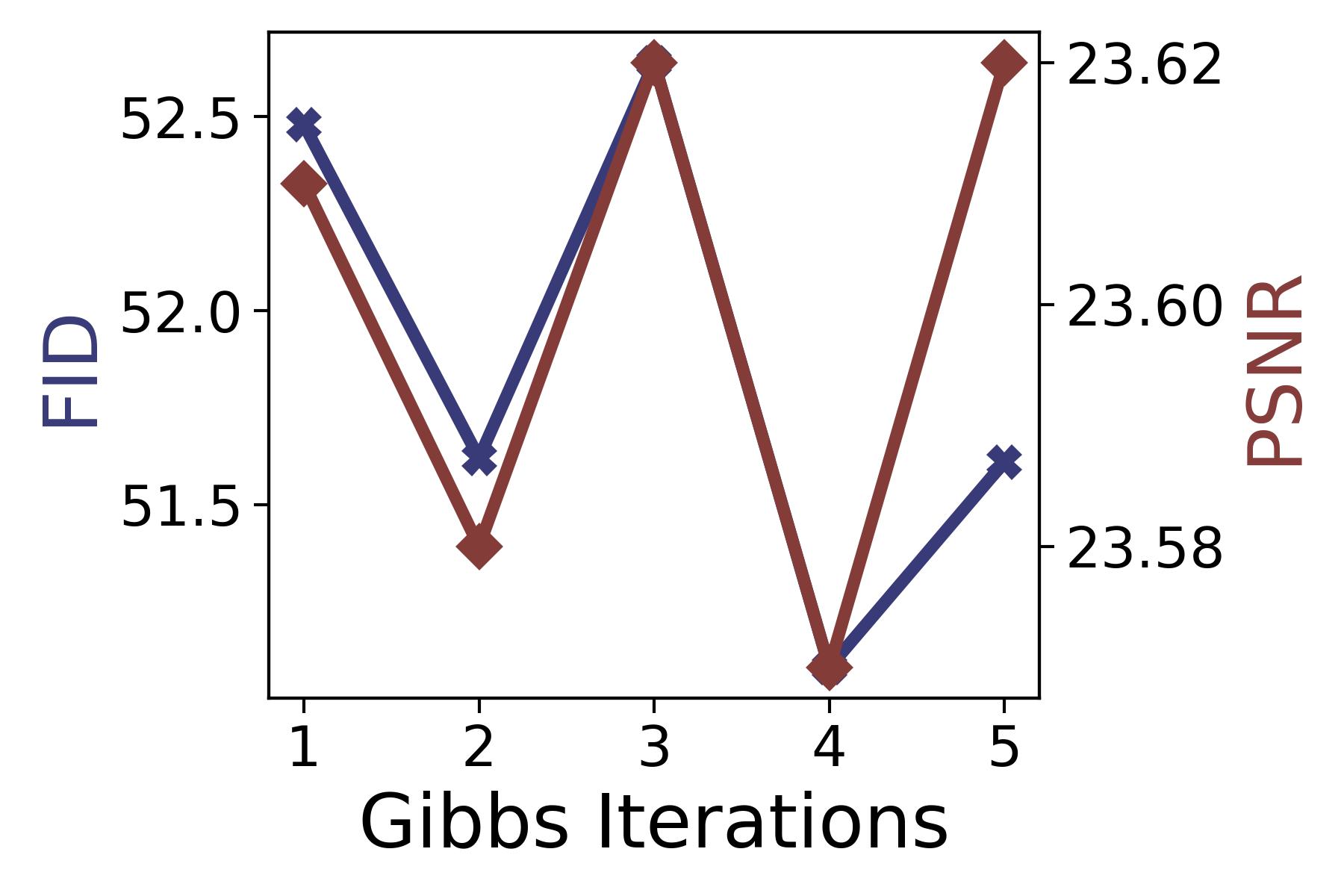}
    \label{fig:gibbs_gd}
    }
     \end{subfigure}
    \caption{FID and PSNR values when increasing the number of Gibbs iterations (with $N=1$ particle).}
    \label{fig:gibbs}
\end{figure}

\section{Proposal Distribution Scaling Coefficients}
\label{sec:scaling_coefs}
Recall that our proposal distribution (\Eqref{eq:proposal} in the main text) is made up of two elements. These elements are scaled by two coefficients, $\gamma_t$ and $\lambda_t$. Here, we provide an explicit formula for these coefficients. We found that our proposed scaling works better than common procedures used in the literature. For consistency with baseline methods, we also used our proposed scaling approach for DPS and TDS (which corresponds to setting $s=0$), since these methods apply a similar update rule. We tried to use it for other baselines but it did not work well for them.

Let $\rvg_t^1 \coloneqq \nabla_{\rvz_t}||\rvy_0 -\gA(\gD(\bar{\rvz}_0(\rvz_t)))||_2^2$, and $\rvg_t^2 \coloneqq \nabla_{\rvmu_\theta(\rvz_{t+1}, t+1)}||\rvy_t - \gA(\gD(\rvmu_\theta(\rvz_{t+1}, t+1)))||_2^2$.
We set the scaling coefficients $\gamma_t$ and $\lambda_t$ according to the following scheme: $\gamma_t = \mathbbm{1}_{[t \geq s]} \cdot \kappa_1 \cdot \frac{1}{max(||\rvg_t^1||_2^2, 1)} + \mathbbm{1}_{[t < s]} \cdot \kappa_2 \cdot (1 - \rho) \cdot \frac{1}{max(||\rvg_t^1||_2^2, 1)}$, and $\lambda_t = \mathbbm{1}_{[t < s]} \cdot \kappa_2 \cdot \rho \cdot \frac{1}{max(||\rvg_t^2||_2^2, 1)}$, where $s, \rho, \kappa_1, \kappa_2$ are all hyper-parameters.

% Here $\{\nu_1, \nu_2\}$ are hyperparameters that controls the effect of the variance scaling, and $\{\kappa_1, \kappa_2\}$ scale the entire term. In practice, since we use either $\rvg_t^1$ or $\rvg_t^2$, but not both, we have only one set of hyperparameters which is used for both $\gamma_t$ and $\lambda_t$. That is, $\nu_1 \coloneqq \nu_2$ and $\kappa_1 \coloneqq \kappa_2$.

% \section{Run Time}
% \label{sec:run_time}

\section{Computational Cost}
\label{sec:computational_cost}
We add here a comparison between the compared methods in average run-time (seconds) and memory (GB) over 10 trials for sampling a single image on ImageNet box inpainting task. For TDS and PFLD we use 5 particles. For \MN{}, we inspect several variants that differ in the number of particles and Gibbs iterations. The results are shown in Table \ref{tab:compute_cost}. From the table, the run-time is roughly linear in the number of particles and Gibbs iterations. Yet, importantly, it can be controlled by the practitioner to trade off performance (which can be good with one Gibbs iteration and one particle) and computational demand. In addition, we note that our code is not properly optimized and improvements can be made to it.

\begin{table*}[!h]
%\small
\centering
\caption{Average run time and memory demand on ImageNet box inpainting task.}
\begin{tabular}{l cc}
    \toprule
    Method & Run Time (Sec.) & Memory (GB) \\
    \midrule
    LDPS & 105.5 & 8.123\\
    LTDS & 418.5 & 19.86\\
    ReSample & 333.4 & 5.769\\
    PSLD & 144.5 & 9.473\\
    PFLD & 469.8 & 26.51\\
    LatentDAPS & 67.82 & 4.745\\
    \midrule
    \MN{} (1 particle) & 136.3 & 9.213\\ 
    \MN{} (3 particles) & 375.1 & 15.11\\
    \MN{} (5 particles) & 537.2 & 21.16\\
    \MN{} (10 particles) & 1013. & 35.78\\
    \MN{} (1 particle; 2 Gibbs iterations) & 271.2 & 9.213\\
    \MN{} (1 particle; 4 Gibbs iterations) & 541.0 & 9.213\\
    \bottomrule
    \end{tabular}
\label{tab:compute_cost}
\end{table*}

\section{Further Ablation Studies}
\label{sec:further_ablations}
\textbf{Number of Gibbs iterations.} Recall that in the main text we used only one Gibbs step throughout. Here we examine the effect of using more Gibbs steps in \Figref{fig:gibbs}. We do so on two tasks, the first is super-resolution and the second in Gaussian deblurring, both on ImageNet. In super-resolution, we observe an improvement in the PSNR with the number of Gibbs steps while suffering from reduction in the FID. Interestingly, this experiment shows that we can obtain a comparable PSNR to ReSample while having a substantially better FID compared to it on this task. In Gaussian deblurring there is a less clear trend, nevertheless, the figure shows that the FID can be improved while maintaining roughly the same PSNR value. We speculate that the difference between these two cases stems from the different $s$ values used in these experiments. Taking $s > 0$ tends to generate smoother images that are better aligned with higher values of distortion metrics such as PSNR.

\textbf{Proposal distributions.} In Table \ref{tab:proposal_distributions} we compare different choices of proposal distributions. Specifically, we compare to (1) the naive choice of using the prior as a proposal distribution, i.e., a bootstrap filter \cite{gordon1993novel}; and (2) taking the DPS step as a proposal distribution, which corresponds to setting $s=0$. The table shows a clear advantage to the proposal distribution used in this study and presented in \Secref{sec:post_sampling}.

\begin{table*}[!h]
%\small
\centering
\caption{\textit{FFHQ}. Proposal distribution ablation. Box inpainting on $1024$ test examples using $N=5$ particles.}
\begin{tabular}{l c cc c ccc}
    \toprule
    && \multicolumn{2}{c}{Perceptual Quality} && \multicolumn{3}{c}{Distortion}\\
    \cmidrule(l){3-4}  \cmidrule(l){6-8}
    && FID ($\downarrow$) & NIQE ($\downarrow$) && PSNR ($\uparrow$) & SSIM ($\uparrow$) & LPIPS ($\downarrow$) \\
    \midrule
    Prior proposal && 62.35 & 7.628 && 12.60 & 0.365 & 0.671\\
    DPS proposal && 39.94 & 7.558 && \textbf{24.20} & 0.814 & 0.235\\
    \midrule
    \MN{} && \textbf{33.87} & \textbf{7.066} && 24.10 & \textbf{0.821} & \textbf{0.211}\\ 
    \bottomrule
    \end{tabular}
\label{tab:proposal_distributions}
\end{table*}

\section{Forward Process}
\begin{figure*}[!t]
    \centering
    \includegraphics[height=1.8\textwidth, width=0.9\textwidth, keepaspectratio]{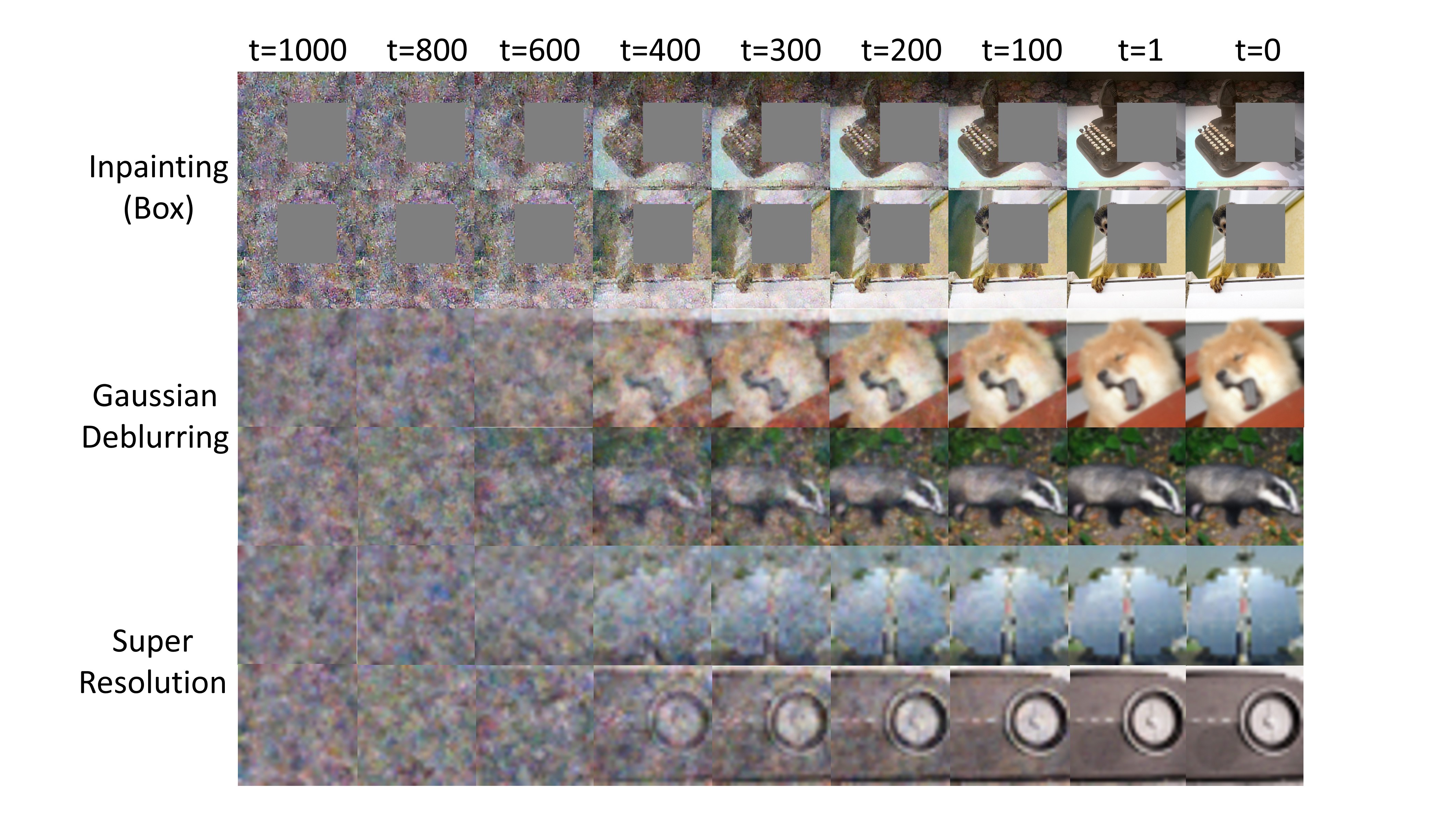}
\caption{Evolution of $\rvy_t$ over time for different tasks according to forward process of DDIM.}
\label{fig:in_forward_process}
\end{figure*}

In \Figref{fig:in_forward_process} we present the evolution of the auxiliary labels $\rvy_t$ over time as part of the forward process according to our proposed sampling procedure in \Secref{sec:sampling_procedure}, steps (a) \& (b). From the figure, we observe a gradual cleaning of noise in the auxiliary labels when advancing from time $t=1000$ to time $t=0$. 

\section{Full Results}
\label{sec:full_results}

\begin{table*}[!h]
\small
\centering
\caption{\textit{ImageNet}. Box inpainting on $1024$ test examples.}
\begin{tabular}{l c cc c ccc}
    \toprule
    && \multicolumn{2}{c}{Perceptual Quality} && \multicolumn{3}{c}{Distortion}\\
    \cmidrule(l){3-4}  \cmidrule(l){6-8}
    && FID ($\downarrow$) & NIQE ($\downarrow$) && PSNR ($\uparrow$) & SSIM ($\uparrow$) & LPIPS ($\downarrow$) \\
    \midrule
    LDPS && 65.04 & 7.935 && \underline{19.51} & 0.665 & 0.379\\
    LTDS && 64.74 & 7.907 && 19.49 & 0.665 & 0.378\\
    ReSample && 90.32 & 8.464 && 18.16 & \underline{0.695} & \textbf{0.318}\\
    PSLD && 71.15 & 8.042 && 18.10 & 0.565 & 0.434\\
    PFLD && 72.83 & 7.933 && 18.09 & 0.564 & 0.436\\
    LatentDAPS && 98.24 & 11.36 && \textbf{19.98} & \textbf{0.704} & 0.394\\
    \midrule
    \MN{} (1 particle) && \underline{51.49} & \textbf{6.878} && 19.38 & 0.672 & 0.326\\ 
    %\MN{} (Ours) && 51.46 & 6.953 && 19.41 & 0.669 & 0.331\\
    \MN{} (5 particles) && \textbf{50.67} & \underline{6.891} && 19.42 & 0.672 & \underline{0.325}\\ 
    \bottomrule
    \end{tabular}
\label{tab:ib_imagenet}
\end{table*}

\begin{table*}[!h]
\small
\centering
\caption{\textit{ImageNet}. Free-form inpainting on $1024$ test examples.}
\begin{tabular}{l c cc c ccc}
    \toprule
    && \multicolumn{2}{c}{Perceptual Quality} && \multicolumn{3}{c}{Distortion}\\
    \cmidrule(l){3-4}  \cmidrule(l){6-8}
    && FID ($\downarrow$) & NIQE ($\downarrow$) && PSNR ($\uparrow$) & SSIM ($\uparrow$) & LPIPS ($\downarrow$) \\
    \midrule
    LDPS && 53.47 & 7.867 && \underline{23.76} & 0.722 & 0.334\\
    LTDS && 52.75 & 7.884 && 23.74 & 0.721 & 0.334\\
    ReSample && 44.15 & 7.104 && 22.78 & \textbf{0.778} & \textbf{0.248}\\
    PSLD && 62.38 & 8.037 && 21.65 & 0.610 & 0.411\\
    PFLD && 63.34 & 8.026 && 21.59 & 0.609 & 0.414\\
    LatentDAPS && 68.65 & 9.625 && \textbf{24.75} & \underline{0.766} & 0.330\\
    \midrule
    \MN{} (1 particle) && \underline{38.21} & \underline{6.969} && 23.54 & 0.727 & 0.285\\ 
    \MN{} (5 particles) && \textbf{36.18} & \textbf{6.671} && 23.43 & 0.727 & \underline{0.278}\\
    \bottomrule
    \end{tabular}
\label{tab:ff_imagenet}
\end{table*}

\begin{table*}[!h]
\small
\centering
\caption{\textit{ImageNet}. Gaussian deblurring on $1024$ test examples.}
\begin{tabular}{l c cc c ccc}
    \toprule
    && \multicolumn{2}{c}{Perceptual Quality} && \multicolumn{3}{c}{Distortion}\\
    \cmidrule(l){3-4}  \cmidrule(l){6-8}
    && FID ($\downarrow$) & NIQE ($\downarrow$) && PSNR ($\uparrow$) & SSIM ($\uparrow$) & LPIPS ($\downarrow$) \\
    \midrule
    LDPS && 52.48 & 6.855 && 23.61 & 0.615 & 0.383\\
    LTDS && \underline{50.82} & \underline{6.695} && 23.57 & 0.614 & \underline{0.379}\\
    ReSample && \textbf{46.45} & 7.411 && \underline{24.36} & \underline{0.639} & \textbf{0.353}\\
    PSLD && 60.68 & \textbf{6.599} && 21.94 & 0.506 & 0.417\\
    PFLD && 60.94 & 6.733 && 21.78 & 0.496 & 0.421\\
    LatentDAPS && 77.09 & 10.55 && \textbf{24.44} & \textbf{0.659} & 0.417\\
    \midrule
    \MN{} (1 particle) && 52.48 & 6.855 && 23.61 & 0.615 & 0.383\\ 
    %\MN{} (5 particles) && 51.95 & 6.796 && 23.60 & 0.614 & 0.382\\ 
    \MN{} (5 particles) && 52.29 & 6.791 && 23.61 & 0.615 & 0.383\\ 
    \bottomrule
    \end{tabular}
\label{tab:gd_imagenet}
\end{table*}

\begin{table*}[!h]
\small
\centering
\caption{\textit{ImageNet}. Super-resolution ($8\times$) on $1024$ test examples.}
\begin{tabular}{l c cc c ccc}
    \toprule
    && \multicolumn{2}{c}{Perceptual Quality} && \multicolumn{3}{c}{Distortion}\\
    \cmidrule(l){3-4}  \cmidrule(l){6-8}
    && FID ($\downarrow$) & NIQE ($\downarrow$) && PSNR ($\uparrow$) & SSIM ($\uparrow$) & LPIPS ($\downarrow$) \\
    \midrule
    LDPS && 61.02 & 6.514 && 21.65 & 0.523 & 0.439\\
    %LTDS && 58.73 & 7.157 && 21.45 & 0.515 & 0.454\\
    LTDS && 59.12 & 6.270 && 21.59 & 0.520 & 0.435\\
    ReSample && 87.65 & 8.290 && \underline{22.05} & \underline{0.532} & 0.491\\
    PSLD && 66.56 & 7.669 && 20.83 & 0.480 & 0.489\\ 
    PFLD && 64.72 & 7.685 && 20.83 & 0.479 & 0.492\\
    LatentDAPS && 104.6 & 12.68 && \textbf{22.38} & \textbf{0.566} & 0.489\\
    \midrule
    \MN{} (1 particle) && \underline{58.06} & \underline{6.243} && 21.51 & 0.524 & \underline{0.434}\\ 
    \MN{} (5 particles) && \textbf{57.89} & \textbf{6.238} && 21.50 & 0.520 & \textbf{0.433}\\ 
    \bottomrule
    \end{tabular}
\label{tab:sr_imagenet}
\end{table*}

\begin{table*}[!h]
\small
\centering
\caption{\textit{FFHQ}. Box inpainting on $1024$ test examples.}
\begin{tabular}{l c cc c ccc}
    \toprule
    && \multicolumn{2}{c}{Perceptual Quality} && \multicolumn{3}{c}{Distortion}\\
    \cmidrule(l){3-4}  \cmidrule(l){6-8}
    && FID ($\downarrow$) & NIQE ($\downarrow$) && PSNR ($\uparrow$) & SSIM ($\uparrow$) & LPIPS ($\downarrow$) \\
    \midrule
    LDPS && 39.81 & 7.592 && 24.15 & 0.814 & 0.236\\
    LTDS && 39.57 & 7.602 && \underline{24.24} & 0.814 & 0.236\\
    ReSample && 86.79 & 7.142 && 19.75 & 0.815 & 0.230\\
    PSLD && 39.68 & \underline{6.544} && 22.31 & 0.774 & 0.246\\
    PFLD && 39.06 & \textbf{6.509} && 22.40 & 0.774 & 0.245\\
    LatentDAPS && 60.24 & 9.999 && \textbf{24.91} & \textbf{0.838} & 0.257\\
    \midrule
    \MN{} (1 particle) && \textbf{33.37} & 7.032 && 23.98 & 0.819 & \underline{0.212}\\ 
    \MN{} (5 particles) && \underline{33.87} & 7.066 && 24.10 & \underline{0.821} & \textbf{0.211}\\ 
    \bottomrule
    \end{tabular}
\label{tab:ib_ffhq}
\end{table*}

\begin{table*}[!h]
\small
\centering
\caption{\textit{FFHQ}. Free-form inpainting on $1024$ test examples.}
\begin{tabular}{l c cc c ccc}
    \toprule
    && \multicolumn{2}{c}{Perceptual Quality} && \multicolumn{3}{c}{Distortion}\\
    \cmidrule(l){3-4}  \cmidrule(l){6-8}
    && FID ($\downarrow$) & NIQE ($\downarrow$) && PSNR ($\uparrow$) & SSIM ($\uparrow$) & LPIPS ($\downarrow$) \\
    \midrule
    LDPS && 40.17 & 7.609 && \underline{27.95} & 0.858 & 0.212\\
    LTDS && 39.78 & 7.578 && \underline{27.95} & 0.858 & 0.212\\
    ReSample && 37.01 & \textbf{6.622} && 26.31 & \textbf{0.891} & \textbf{0.151}\\
    PSLD && 36.26 & 6.835 && 26.19 & 0.823 & 0.216\\ 
    PFLD && 36.43 & \underline{6.817} && 26.35 & 0.825 & 0.215\\
    LatentDAPS && 54.40 & 8.766 && \textbf{29.23} & \underline{0.883} & 0.223\\
    \midrule
    \MN{} (1 particle) && \underline{33.67} & 7.034 && 27.35 & 0.859 & \underline{0.194}\\ 
    \MN{} (5 particles) && \textbf{33.60} & 7.021 && 27.35 & 0.859 & \underline{0.194}\\
    \bottomrule
    \end{tabular}
\label{tab:ff_ffhq}
\end{table*}

\begin{table*}[!h]
\small
\centering
\caption{\textit{FFHQ}. Gaussian deblurring on $1024$ test examples.}
\begin{tabular}{l c cc c ccc}
    \toprule
    && \multicolumn{2}{c}{Perceptual Quality} && \multicolumn{3}{c}{Distortion}\\
    \cmidrule(l){3-4}  \cmidrule(l){6-8}
    && FID ($\downarrow$) & NIQE ($\downarrow$) && PSNR ($\uparrow$) & SSIM ($\uparrow$) & LPIPS ($\downarrow$) \\
    \midrule
    LDPS && \underline{29.30} & \underline{6.538} && 28.03 & 0.775 & 0.237\\
    LTDS && 30.23 & 6.553 && 27.93 & 0.772 & 0.238\\
    ReSample && 39.80 & 7.441 && \underline{28.45} & 0.763 & 0.275\\
    PSLD && 36.31 & 6.802 && 24.02 & 0.633 & 0.341\\
    PFLD && 37.16 & 6.751 && 23.96 & 0.628 & 0.343\\
    LatentDAPS && 54.28 & 9.496 && \textbf{29.70} & \textbf{0.831} & 0.283\\
    \midrule
    \MN{} (1 particle) && \textbf{29.19} & 6.575 && 28.37 & \underline{0.789} & \textbf{0.232}\\ 
    \MN{} (5 particles) && 29.47 & \textbf{6.528} && 28.34 & 0.787 & \underline{0.233}\\ 
    \bottomrule
    \end{tabular}
\label{tab:gd_ffhq}
\end{table*}

\begin{table*}[!h]
\small
\centering
\caption{\textit{FFHQ}. Super-resolution ($8 \times$) on $1024$ test examples.}
\begin{tabular}{l c cc c ccc}
    \toprule
    && \multicolumn{2}{c}{Perceptual Quality} && \multicolumn{3}{c}{Distortion}\\
    \cmidrule(l){3-4}  \cmidrule(l){6-8}
    && FID ($\downarrow$) & NIQE ($\downarrow$) && PSNR ($\uparrow$) & SSIM ($\uparrow$) & LPIPS ($\downarrow$) \\
    \midrule
    LDPS && \textbf{29.64} & \textbf{6.412} && 25.48 & 0.701 & 0.282\\
    LTDS && 30.45 & \textbf{6.412} && 25.38 & 0.698 & 0.284\\
    ReSample && 59.23 & 7.307 && \underline{25.55} & 0.661 & 0.356\\
    PSLD && 40.33 & 6.803 && 23.66 & 0.615 & 0.347\\
    PFLD && 38.11 & 6.832 && 23.69 & 0.617 & 0.345\\
    LatentDAPS && 70.24 & 10.17 && \textbf{26.87} & \textbf{0.760} & 0.344\\
    \midrule
    \MN{} (1 particle) && \underline{30.02} & 6.426 && 25.52 & \underline{0.706} & \textbf{0.277}\\ 
    %\MN{} (Ours) && 30.07 & 6.362 && 25.46 & 0.703 & 0.280\\ 
    \MN{} (5 particles) && 30.62 & 6.455 && 25.49 & 0.703 & \underline{0.278}\\ 
    \bottomrule
    \end{tabular}
\label{tab:sr_ffhq}
\end{table*}

\clearpage
\section{Additional Image Reconstructions}
\label{sec:img_rec}

\begin{figure*}[!h]
    \centering
    \includegraphics[height=1.1\textwidth, width=0.9\textwidth, keepaspectratio]{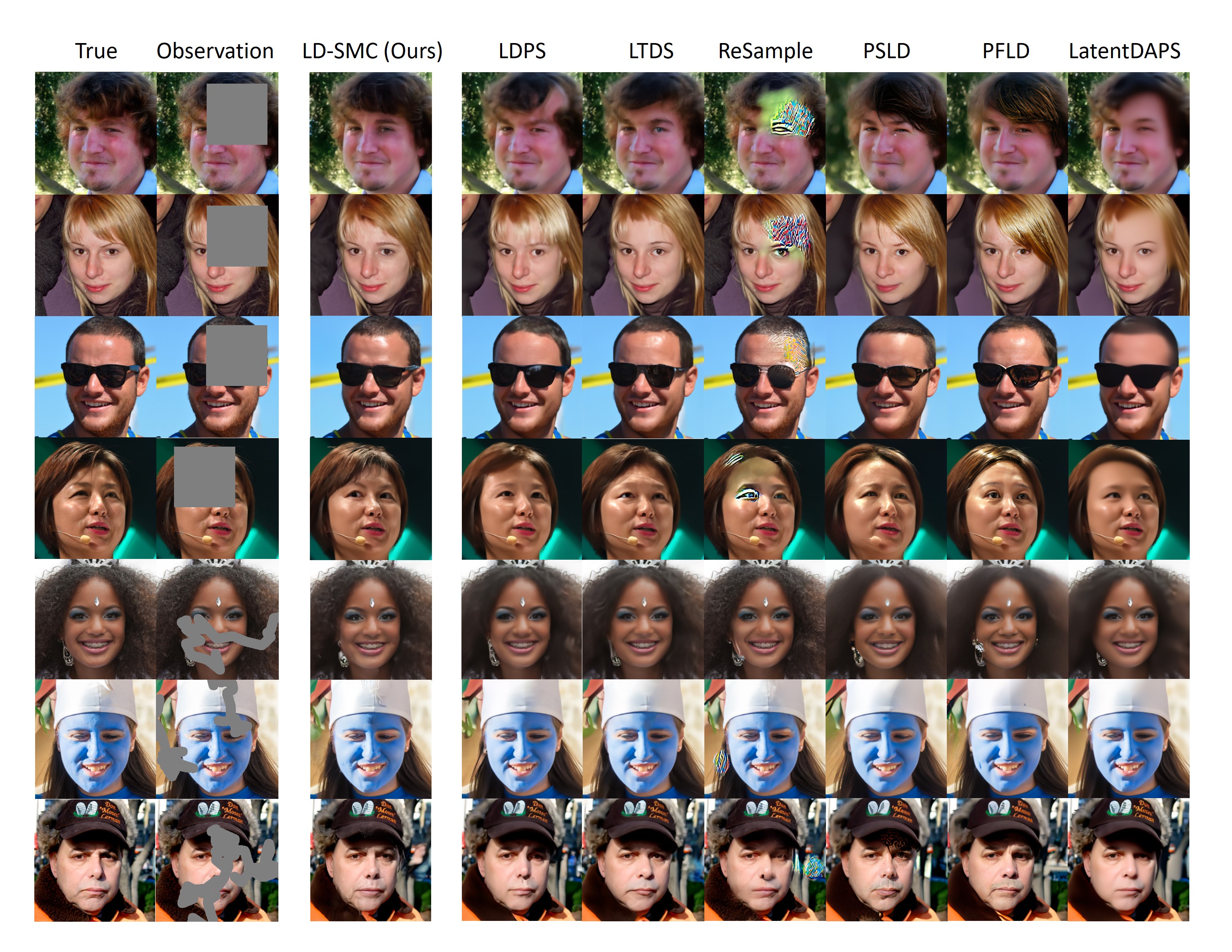}
\caption{Comparison between \MN{} and baseline methods on inpainting of FFHQ images.}
\label{fig:ffhq_ib_recon}
\end{figure*}

\begin{figure*}[!h]
    \centering
    \includegraphics[height=1.1\textwidth, keepaspectratio]{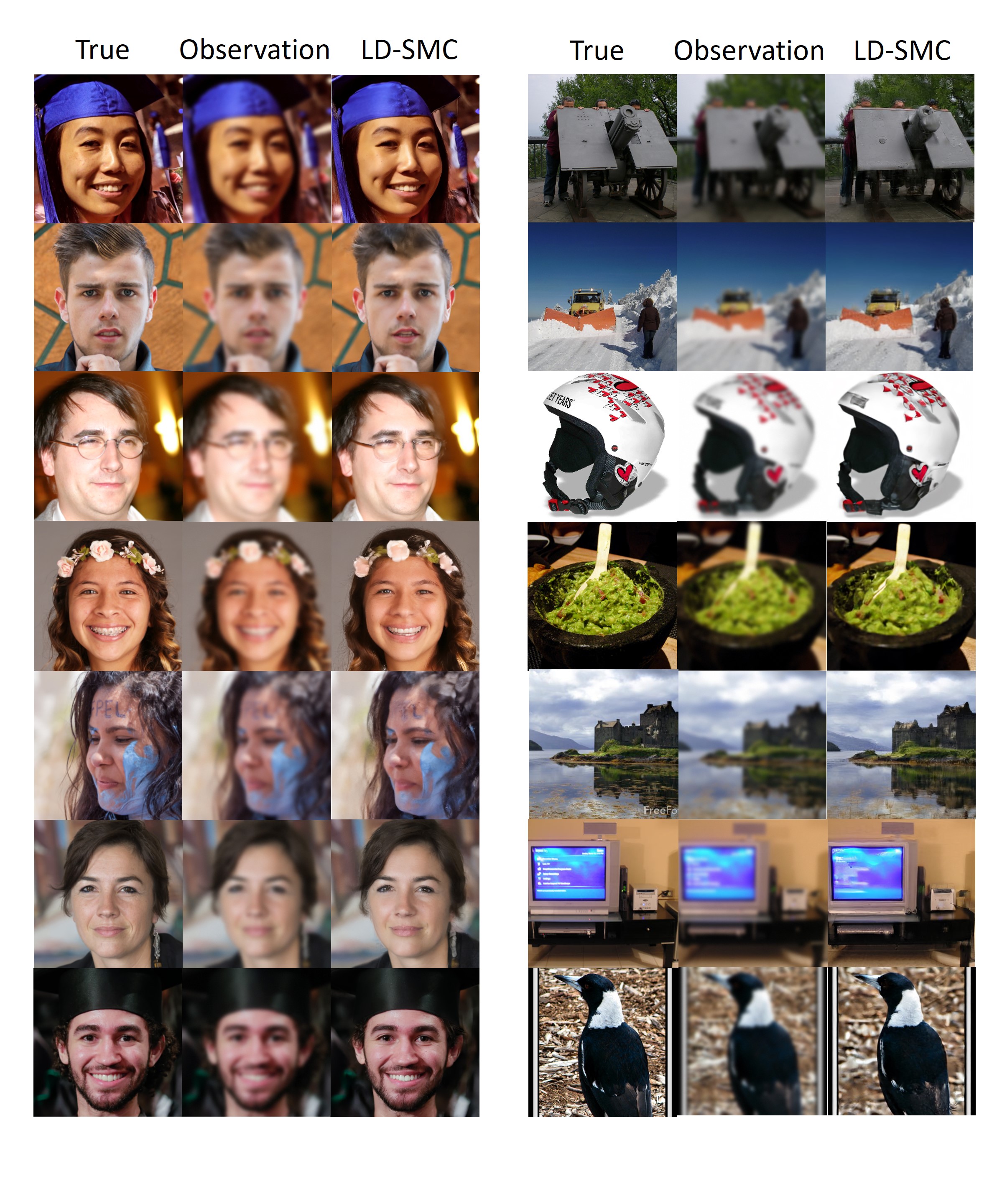}
\caption{\textit{Gaussian deblurring}. \MN{} reconstruction of images from FFHQ (left) and ImageNet (right).}
\label{fig:gd_recon}
\end{figure*}

\begin{figure*}[!h]
    \centering
    \includegraphics[height=1.1\textwidth, keepaspectratio]{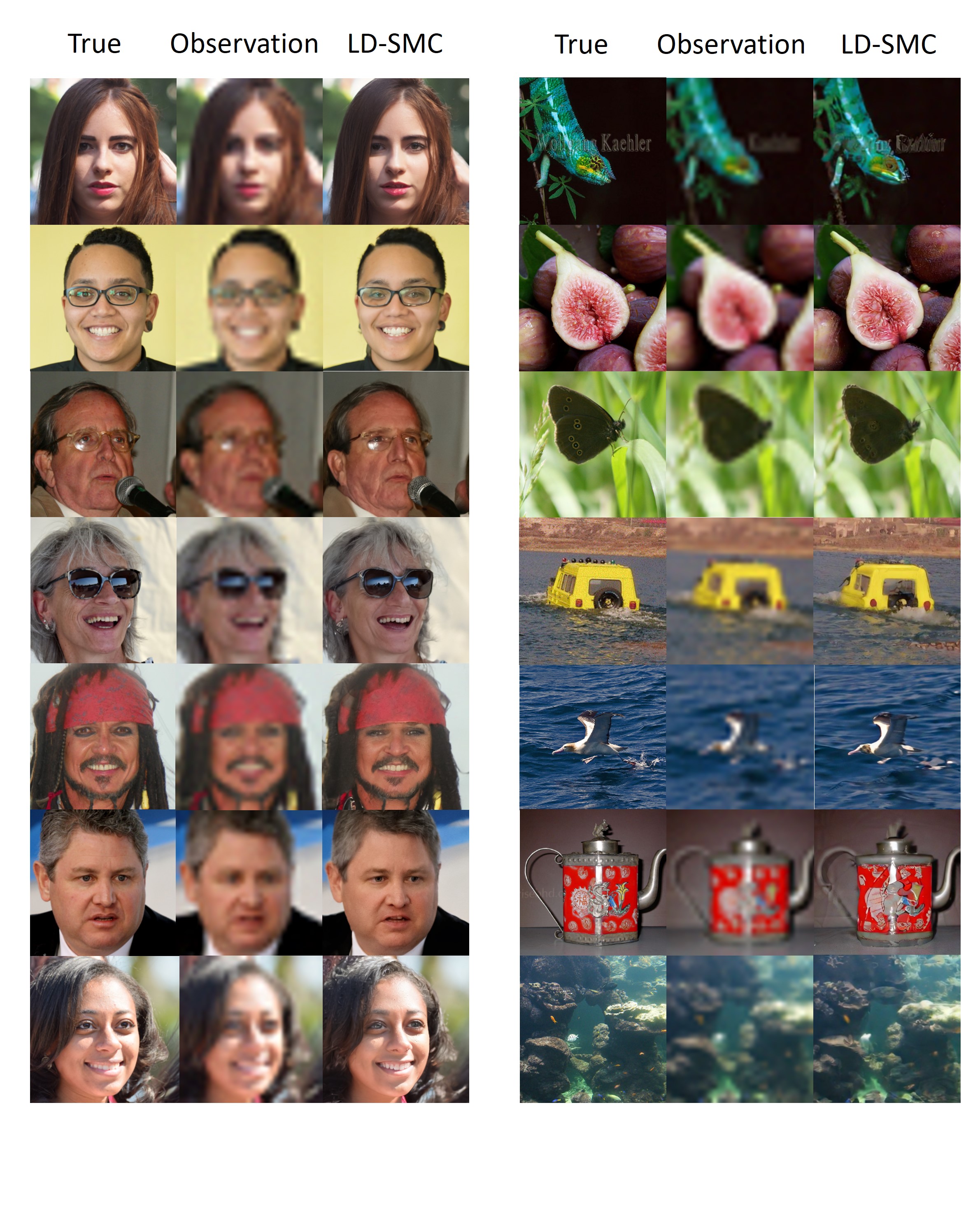}
\caption{\textit{Super-resolution}. \MN{} reconstruction of images from FFHQ (left) and ImageNet (right).}
\label{fig:sr_recon}
\end{figure*}

% \begin{figure*}[!t]
%     \centering
%     \includegraphics[height=1.1\textwidth, keepaspectratio]{figures/ffhq_deblurring.jpg}
% \caption{\textit{Gaussian deblurring}. FFHQ image reconstructions by \MN{}.}
% \label{fig:ffhq_gd_recon}
% \end{figure*}

% \begin{figure*}[!t]
%     \centering
%     \includegraphics[height=1.1\textwidth, keepaspectratio]{figures/ffhq_super_res.jpg}
% \caption{\textit{Super-Resolution}. FFHQ image reconstructions by \MN{}.}
% \label{fig:ffhq_sr_recon}
% \end{figure*}

% \begin{figure*}[!t]
%     \centering
%     \includegraphics[height=1.1\textwidth, keepaspectratio]{figures/imagenet_deblurring.jpg}
% \caption{\textit{Gaussian deblurring}. ImageNet image reconstructions by \MN{}.}
% \label{fig:in_gd_recon}
% \end{figure*}

% \begin{figure*}[!t]
%     \centering
%     \includegraphics[height=1.1\textwidth, keepaspectratio]{figures/imagenet_super_res.jpg}
% \caption{\textit{Super-Resolution}. ImageNet image reconstructions by \MN{}.}
% \label{fig:in_sr_recon}
% \end{figure*}
% \input{Asymptotic_proof}
% \input{rebuttle_tables}
% \appendix
% \onecolumn
% \section{You \emph{can} have an appendix here.}

%%%%%%%%%%%%%%%%%%%%%%%%%%%%%%%%%%%%%%%%%%%%%%%%%%%%%%%%%%%%%%%%%%%%%%%%%%%%%%%
%%%%%%%%%%%%%%%%%%%%%%%%%%%%%%%%%%%%%%%%%%%%%%%%%%%%%%%%%%%%%%%%%%%%%%%%%%%%%%%

\end{document}